\documentclass[sigconf]{acmart}

\usepackage{algorithm}
\usepackage{algorithmic}
\usepackage{amsmath}
\usepackage{booktabs}
\usepackage{nicefrac}
\usepackage{microtype}
\usepackage{xcolor}
\usepackage{subcaption}
\usepackage{multirow}
\usepackage{makecell}
\usepackage{pifont}
\usepackage{array}
\usepackage{enumitem}
\usepackage{colortbl}
\newcolumntype{P}[1]{>{\raggedright\arraybackslash}p{#1}}
\definecolor{yellowtext}{RGB}{68,132,243}
\definecolor{yellowred}{RGB}{50,167,82}
\definecolor{yellowblue}{RGB}{251,191,5}
\definecolor{darkblue}{rgb}{0, 0, 0.5}
\definecolor{darkgreen}{rgb}{0.0, 0.42, 0.24}
\definecolor{maroon}{HTML}{A00000}
\definecolor{gray}{rgb}{0.5, 0.5, 0.5}
\definecolor{chocolate}{HTML}{D2691E}
\definecolor{indigo}{HTML}{4B0082}
\definecolor{violet}{HTML}{4B2E83}
\definecolor{lightgreen}{HTML}{E0FBE0}
\definecolor{lightred}{HTML}{FBE0E0}
\definecolor{paperorange}{HTML}{C04000}
\definecolor{lightblue}{rgb}{0.0, 0.0, 0.5}
\definecolor{cadmiumgreen}{rgb}{0.0, 0.42, 0.24}
\definecolor{forestgreen}{rgb}{0.13, 0.55, 0.13}
\definecolor{lightbluebg}{RGB}{235, 245, 255}
\definecolor{blueframe}{RGB}{70, 130, 180}
\definecolor{DeepPurple}{RGB}{103, 58, 183}
\definecolor{LightPurple}{RGB}{237, 231, 246}
\definecolor{lightgray}{rgb}{0.9, 0.9, 0.9}

\newcommand{\TextCircle}[1][0.7]{%
    \textcolor{yellowtext}{\textbf{[T]}}%
}
\newcommand{\ImageCircle}[1][0.76]{%
    \textcolor{yellowred}{\textbf{[I]}}%
}

\newcommand{\stddev}[1]{\textcolor{gray}{\scalebox{.8}{$\pm$#1}}}

\newcommand{\singlechoice}{\ensuremath{\bigcirc}}
\newcommand{\multichoice}{\ensuremath{\square}}

\newcommand{\circlednum}[1]{%
  \textcolor{DeepPurple}{\textbf{(#1)}}%
}

\newsavebox{\myboxcontent}
\newenvironment{custommdframed}
{%
  \par\vspace{8pt}\noindent
  \setlength{\fboxsep}{6pt}%
  \begin{lrbox}{\myboxcontent}%
    \begin{minipage}{\dimexpr\linewidth-2\fboxsep\relax}%
      \setlength{\parindent}{0pt}
      \ignorespaces
}{%
    \end{minipage}%
  \end{lrbox}%
  \colorbox{LightPurple!40}{\usebox{\myboxcontent}}%
  \par\vspace{8pt}%
}

\newenvironment{PromptBox}[1][]
  {%
   \par\vspace{0.5em}\noindent
   \setlength{\fboxsep}{6pt}%
   \begin{lrbox}{\myboxcontent}%
   \begin{minipage}{\dimexpr\linewidth-12pt-2\fboxrule\relax}%
   \small\ttfamily
  }
  {%
   \end{minipage}%
   \end{lrbox}%
   \fcolorbox{black!20}{gray!3}{\usebox{\myboxcontent}}%
   \par\vspace{0.5em}%
  }

\AtBeginDocument{%
  }

\begin{document}

\copyrightyear{2026}
\acmYear{2026}
\setcopyright{cc}
\setcctype{by}
\acmConference[CHI '26]{Proceedings of the 2026 CHI Conference on Human Factors in Computing Systems}{April 13--17, 2026}{Barcelona, Spain}
\acmBooktitle{Proceedings of the 2026 CHI Conference on Human Factors in Computing Systems (CHI '26), April 13--17, 2026, Barcelona, Spain}
\acmPrice{}
\acmDOI{10.1145/3772318.3791458}
\acmISBN{979-8-4007-2278-3/2026/04}

\acmSubmissionID{7676}



\settopmatter{authorsperrow=4}

% 1. Yinghao Zhu (Corresponding Author, 2 Affiliations)
\author{Yinghao Zhu}
\authornote{Equal contribution.}
\orcid{0000-0002-2640-6477}
\affiliation{%
  \institution{Peking University}
  % \department{National Engineering Research Center for Software Engineering}
  \city{Beijing}
  \country{China}
}
\affiliation{%
  \institution{University of Hong Kong}
  % \department{School of Computing and Data Science}
  \city{Hong Kong}
  \country{China}
}
\email{yhzhu99@gmail.com}

% 2. Dehao Sui
\author{Dehao Sui}
\authornotemark[1]
\orcid{0009-0000-9081-059X}
\affiliation{%
  \institution{Peking University}
  % \department{National Engineering Research Center for Software Engineering}
  \city{Beijing}
  \country{China}
}
\email{dehaosui1@gmail.com}

% 3. Zixiang Wang
\author{Zixiang Wang}
\authornotemark[1]
\orcid{0009-0000-1257-9580}
\affiliation{%
  \institution{Peking University}
  % \department{National Engineering Research Center for Software Engineering}
  \city{Beijing}
  \country{China}
}
\email{wangzx@stu.pku.edu.cn}

% 4. Xuning Hu
\author{Xuning Hu}
\orcid{0009-0009-1305-2081}
\affiliation{%
  \institution{Hong Kong University of Science and Technology}
  % \department{Department of Computer Science and Engineering}
  \city{Hong Kong}
  \country{China}
}
\email{xuninghu@hkust-gz.edu.cn}

% 5. Lei Gu
\author{Lei Gu}
\orcid{0009-0000-9144-1677}
\affiliation{%
  \institution{Peking University}
  % \department{National Engineering Research Center for Software Engineering}
  \city{Beijing}
  \country{China}
}
\email{leiguha99@gmail.com}

% 6. Yifan Qi
\author{Yifan Qi}
\orcid{0009-0008-5740-2832}
\affiliation{%
  \institution{Peking University}
  % \department{National Engineering Research Center for Software Engineering}
  \city{Beijing}
  \country{China}
}
\email{qiyifan2004@icloud.com}

% 7. Tianchen Wu
\author{Tianchen Wu}
\orcid{0000-0003-2974-7928}
\affiliation{%
  \institution{Peking University Third Hospital}
  % \department{Department of Obstetrics and Gynecology}
  \city{Beijing}
  \country{China}
}
\email{wtc17@pku.edu.cn}

% 8. Ling Wang
\author{Ling Wang}
\orcid{0009-0008-4653-7184}
\affiliation{%
  \institution{Affiliated Xuzhou Municipal Hospital of Xuzhou Medical University}
  % \department{Department of Nephrology}
  \city{Jiangsu}
  \country{China}
}
\email{lingw1228@sohu.com}

% 9. Yuan Wei
\author{Yuan Wei}
\orcid{0000-0003-3387-7549}
\affiliation{%
  \institution{Peking University Third Hospital}
  % \department{Department of Obstetrics and Gynecology}
  \city{Beijing}
  \country{China}
}
\email{weiyuanbysy@163.com}

% 10. Wen Tang
\author{Wen Tang}
\orcid{0000-0002-2263-2979}
\affiliation{%
  \institution{Peking University Third Hospital}
  % \department{Department of Nephrology}
  \city{Beijing}
  \country{China}
}
\email{tanggwen@126.com}

% 11. Zhihan Cui
\author{Zhihan Cui}
\orcid{0000-0003-0883-2322}
\affiliation{%
  \institution{Peking University}
  % \department{Institute for Global Health and Development}
  \city{Beijing}
  \country{China}
}
\email{zhihancui@pku.edu.cn}

% 12. Yasha Wang
\author{Yasha Wang}
\orcid{0000-0002-8026-9688}
\affiliation{%
  \institution{Peking University}
  % \department{National Engineering Research Center for Software Engineering}
  \city{Beijing}
  \country{China}
}
\email{wangyasha@pku.edu.cn}

% 13. Lequan Yu
\author{Lequan Yu}
\orcid{0000-0002-9315-6527}
\affiliation{%
  \institution{University of Hong Kong}
  % \department{School of Computing and Data Science}
  \city{Hong Kong}
  \country{China}
}
\email{lqyu@hku.hk}

% 14. Ewen M Harrison
\author{Ewen M Harrison}
\orcid{0000-0002-5018-3066}
\affiliation{%
  \institution{University of Edinburgh}
  % \department{Centre for Medical Informatics}
  \city{Edinburgh}
  \country{United Kingdom}
}
\email{ewen.harrison@ed.ac.uk}

% 15. Junyi Gao (2 Affiliations)
\author{Junyi Gao}
\orcid{0000-0002-4951-8682}
\affiliation{%
  \institution{University of Edinburgh}
  % \department{Centre for Medical Informatics}
  \city{Edinburgh}
  \country{United Kingdom}
}
\affiliation{%
  \institution{Health Data Research UK}
  \city{London}
  \country{United Kingdom}
}
\email{junyii.gao@gmail.com}

% 16. Liantao Ma
\author{Liantao Ma}
\authornote{Corresponding author.}
\orcid{0000-0001-5233-0624}
\affiliation{%
  \institution{Peking University}
  % \department{National Engineering Research Center for Software Engineering}
  \city{Beijing}
  \country{China}
}
\email{malt@pku.edu.cn}

\renewcommand{\shortauthors}{Zhu et al.}

\title[Augmenting Clinical Decision-Making with an Interactive and Interpretable AI Copilot]{Augmenting Clinical Decision-Making with an Interactive and Interpretable AI Copilot: A Real-World User Study with Clinicians in Nephrology and Obstetrics}

\begin{abstract}
Clinician skepticism toward opaque AI hinders adoption in high-stakes healthcare. We present AICare, an interactive and interpretable AI copilot for collaborative clinical decision-making. By analyzing longitudinal electronic health records, AICare grounds dynamic risk predictions in scrutable visualizations and LLM-driven diagnostic recommendations. Through a within-subjects counterbalanced study with 16 clinicians across nephrology and obstetrics, we comprehensively evaluated AICare using objective measures (task completion time and error rate), subjective assessments (NASA-TLX, SUS, and confidence ratings), and semi-structured interviews. Our findings indicate AICare's reduced cognitive workload. Beyond performance metrics, qualitative analysis reveals that trust is actively constructed through verification, with interaction strategies diverging by expertise: junior clinicians used the system as cognitive scaffolding to structure their analysis, while experts engaged in adversarial verification to challenge the AI's logic. This work offers design implications for creating AI systems that function as transparent partners, accommodating diverse reasoning styles to augment rather than replace clinical judgment.
\end{abstract}


\maketitle

\section{Introduction}

The integration of Artificial Intelligence (AI) into the medical domain is currently defined by a sharp contrast. On one hand, algorithmic performance continues to break records in controlled experiments~\cite{topol2019high, rajkomar2019machine, esteva2019guide,zhu2024emerge,zhu2024prism}. Recent Large Language Models (LLMs) and deep learning architectures often match or exceed human experts in diagnostic tasks ranging from radiology to medical licensing examinations~\cite{singhal2023large, gao2024comprehensive, rajashekar2024human}. On the other hand, the successful and sustained deployment of these systems into actual clinical practice remains rare~\cite{beede2020human, asan2020artificial, zajkac2025towards}. This disconnect is widely termed the ``last mile'' problem of medical AI~\cite{osman2021realizing, thieme2025challenges}. It suggests that the primary barrier to adoption is no longer strictly computational, but related to clinical integration. Current failures in deployment often stem not from a lack of accuracy, but from a lack of clinical legitimacy, specifically the degree to which AI systems align with the complex and accountability-driven reasoning processes that define high-stakes medical practice~\cite{tonekaboni2019clinicians, burgess2023healthcare, zajkac2025towards}.

Early paradigms in Clinical Decision Support Systems (CDSS) typically conceptualized AI as an authoritative predictor. These systems function as black boxes that deliver a final verdict, such as a risk score or a binary diagnosis, and expect passive acceptance from the human user~\cite{zhang2024rethinking, hussain2024development}. However, clinical decision-making is rarely a binary choice that can be automated. Instead, it is an iterative process involving hypothesis generation, evidence gathering, and cross-checking, where clinicians must weigh conflicting data points against patient history and clinical guidelines~\cite{sox2013medical, thieme2025challenges, zhang2024rethinking}. When AI systems bypass these intermediate cognitive steps to offer only a final conclusion, they may act as competitors to the clinician rather than supporters. This dynamic can trigger skepticism or resistance among experts who feel their professional agency is threatened~\cite{zhang2024rethinking, bach2023if, calisto2023assertiveness}. Furthermore, the lack of transparency prevents clinicians from performing their ethical duty of verification, turning the clinician into a liable agent who is responsible for AI errors without possessing the means to interrogate or understand them~\cite{hussain2024development, bienefeld2024human}.

To bridge this gap between algorithms and practice, Human-Computer Interaction (HCI) research must pivot from designing automated predictors to designing AI copilots. These are collaborative partners designed not to dictate decisions, but to support the user's inquiry and process integration~\cite{thieme2025challenges, yang2019unremarkable, yildirim2024multimodal}. Unlike a static predictor, an AI copilot facilitates clinical legitimacy not by demanding blind trust, but by providing the interactive mechanisms for clinicians to triangulate evidence and actively interrogate the underlying data~\cite{cai2019hello, mcgrath2025collaborative, thieme2023designing}.

While the HCI community has begun to explore these themes, existing work remains limited by two critical factors. First, much of the current literature relies on formative interviews or the evaluation of non-interactive, static prototypes. There is a scarcity of studies that empirically evaluate how clinicians interact with fully functional systems capable of real-time data exploration in high-stakes environments. Second, prior work often treats clinicians as a monolithic group. This overlooks how the ``last mile'' challenges may diverge based on domain expertise. While theoretical frameworks warn of de-skilling for junior clinicians or algorithmic aversion from experts, there is limited empirical evidence demonstrating how interactive features might mitigate these risks by supporting distinct cognitive strategies for novices versus experts within the same system~\cite{bienefeld2024human, gaube2021ai, wolf2025clinical}.

In this work, we present and evaluate \textbf{AICare}, an interactive and interpretable AI copilot. AICare translates the concept of clinical legitimacy by exposing its reasoning through dynamic risk trajectories, interactive feature-level attribution, and LLM-synthesized clinical narratives. Rather than delivering a static prediction, AICare grounds its dynamic risk predictions in interpretable visualizations, allowing the clinician to analyze the patient case collaboratively. AICare features the following core components designed to support clinical sensemaking:
\begin{enumerate}[leftmargin=*, label=(\arabic*), topsep=0pt]
    \item A \textbf{dynamic risk trajectory visualization} charting a patient's risk over time, providing a longitudinal narrative of their health journey.
    \item An \textbf{interactive list of critical risk factors} showing the clinical features most influential to a prediction, with on-demand drill-down into specific feature trends.
    \item An \textbf{LLM-driven diagnostic recommendation} that synthesizes the AI's key findings into a concise, clinical narrative.
    \item A \textbf{population-level indicator analysis} that contextualizes a patient's data by comparing their indicators against cohort-level trends.
\end{enumerate}

To assess AICare's impact in a real-world clinical setting, we conducted a within-subjects user study with 16 clinicians utilizing the deployed system. We contend that one of the most complex and critical challenges for clinical AI is the management of chronic conditions through long-term, longitudinal patient follow-up~\cite{choi2016doctor, shickel2017deep}, which involves reasoning over sparse, high-dimensional, and irregularly sampled time-series data~\cite{ma2020concare}. Our preliminary stakeholder discussions further prioritized these high-risk contexts, citing the urgent need for support where error tolerance is minimal and the cognitive load of monitoring deterioration is highest. Therefore, we purposefully selected two representative specialties for our evaluation: nephrology, focusing on chronic disease management for patients with end-stage renal disease, and obstetrics, focusing on preterm birth risk assessment during prenatal care. Both scenarios involve high-stakes decisions where AI assistance could prevent severe adverse outcomes.

To guide our evaluation of AICare within these contexts and to unpack the resulting behavioral dynamics, we structure our investigation around the following three Research Questions (RQs):
\begin{itemize}[leftmargin=*, topsep=0pt]
    \item \textbf{RQ1:} How do clinicians perceive the utility and usability of AICare's interactive and interpretable modules when integrated into their diagnostic workflow?
    \item \textbf{RQ2:} What is the impact of AICare on clinicians' diagnostic efficiency, accuracy, and cognitive workload, and how does this impact vary across different clinical contexts and experience levels?
    \item \textbf{RQ3:} How do the interactive and interpretable features of AICare influence clinicians' trust in the AI's recommendations and their overall decision-making confidence?
\end{itemize}

Through task-based sessions and semi-structured interviews, our findings reveal that clinicians engaged with AICare not as an infallible authority, but as a cognitive partner for collaborative sensemaking. Quantitatively, we observed a significant reduction in perceived cognitive workload ($p=.023$) and a significant increase in diagnostic confidence ($p=.018$), while diagnostic accuracy remained comparable to the baseline. Regarding efficiency, although total task duration did not differ statistically, senior clinicians took slightly longer to complete tasks. This temporal trend aligns with granular interaction metrics that revealed a significant behavioral divergence ($p<.05$) based on expertise: senior clinicians engaged in more active data interrogation behaviors, acting out a process of adversarial verification. Conversely, junior clinicians exhibited lower interaction frequencies, leveraging the system primarily as a cognitive scaffold to structure their analysis.

Qualitatively, we found that trust was not passively granted but actively constructed through a process of human-AI alignment. Clinicians consistently used AICare's interactive features, such as tracing the patient's story along the risk trajectory and scrutinizing the feature importance list, to cross-reference the AI's rationale against their own mental model. This transformed the interaction from a simple consumption of a prediction into a collaborative dialogue. Crucially, moments of disagreement, where the AI's reasoning diverged from clinical intuition, did not necessarily erode trust. Instead, when supported by plausible evidence, they often became opportunities for critical reflection.

Furthermore, clinicians articulated a clear vision for the future of such tools. Beyond its current analytical capabilities, they expressed a strong desire for AI copilots that provide more context-aware and actionable recommendations, such as suggesting specific follow-up tests or linking counter-intuitive findings to relevant clinical guidelines. They envision a future where these systems are seamlessly integrated into the EHR, functioning as vigilant assistants that help them manage information overload to augment rather than replace their irreplaceable clinical judgment.

Overall, this paper makes the following contributions:
\begin{itemize}[leftmargin=*, topsep=0pt]
    \item The design and implementation of AICare, an interactive and interpretable AI copilot currently integrated into hospital information systems and deployed in clinical settings (accessible for demonstration at \url{https://aicare.pages.dev/}), translating a collaborative human-AI paradigm through scrutable, multi-level explanations.
    \item An empirical evaluation combining quantitative metrics and qualitative analysis with practicing clinicians in two distinct specialties and hospital tiers, providing insights into both shared principles and context-specific needs for AI adoption.
    \item A set of actionable design implications for creating AI copilots that foster a collaborative human-AI partnership, grounded in empirical evidence of distinct verification mechanisms: cognitive scaffolding for novices and adversarial verification for experts.
\end{itemize}

\section{Related Work}

Our research synthesizes three critical discourses in HCI and medical AI: the shift from static prediction to collaborative workflow integration, the reconceptualization of trust as an active verification process, and the mediating roles of clinical expertise and context in human-AI interaction.

\subsection{Human-AI Collaboration in Clinical Workflows}
The prevailing challenge in medical AI is no longer strictly regarding algorithmic capability, but rather the sociotechnical gap or ``last mile'' problem of deployment~\cite{osman2021realizing, zajkac2025towards, beede2020human}. Historically, Clinical Decision Support Systems (CDSS) often functioned as rule-based alert systems, providing reminders or flagging potential drug interactions~\cite{sutton2020overview, berner2016clinical, miller1994medical}. With the advent of machine learning, these systems evolved to offer predictive analytics, such as risk scores for patient deterioration or disease onset~\cite{shickel2018deep}, with prominent examples like the Epic Sepsis Model providing a single risk score to alert clinicians~\cite{wong2021external}.

However, recent scholarship argues that this model of AI as a diagnostic authority fundamentally misunderstands the nature of clinical work. Medical decision-making is inherently collaborative, iterative, and socially situated~\cite{yang2019unremarkable, zhang2024rethinking, thieme2025challenges}. Yang et al. demonstrate that successful AI integration requires systems to be ``unremarkable'', implying they must be embedded seamlessly into the social fabric of decision-making rather than acting as disruptive external authorities that demand attention~\cite{yang2019unremarkable}. Similarly, ethnographic work has revealed that high-performing algorithms often fail because they lack the flexibility to handle the environmental realities of clinical workflows, such as poor lighting, internet latency, or patient hardship~\cite{beede2020human, wang2021brilliant}. These failures occur because rigid systems act as competitors to the clinician, challenging the authority of the clinician without understanding the full patient context~\cite{zhang2024rethinking, thieme2025challenges}.

To address this misalignment, the field is moving toward human-AI teaming and copilot models. Zhang et al. propose systems that support the intermediate stages of decision-making, such as hypothesis generation and data gathering, rather than attempting to automate the final verdict~\cite{zhang2024rethinking}. This aligns with findings that clinical usefulness is distinct from algorithmic accuracy; usefulness depends on the configurability of the system to local workflows, such as prioritizing worklists or drafting reports, rather than just diagnosing pathology~\cite{zajkac2025towards, yildirim2024multimodal, thieme2025challenges}. Furthermore, Hussain et al. argue that unless interaction models shift from passive recommendation to active data exploration (e.g., exposing underlying data), clinicians become ``liability sinks'', responsible for AI errors they cannot understand or prevent because the system is inscrutable~\cite{hussain2024development}. Our work answers these calls by evaluating a system designed specifically to support these intermediate reasoning steps, allowing clinicians to retain agency while leveraging AI for complex longitudinal data synthesis.

\subsection{Interpretability and Trust Calibration in Medical AI}
Traditionally, trust in medical AI has been modeled as a static attitude or a reliance score calibrated by the observed accuracy of the model~\cite{lee2004trust}. However, recent literature frames trust as a dynamic, emergent state that is maintained through continuous interaction and verification~\cite{mcgrath2025collaborative, tun2025trust}. Yang et al. further attribute this difficulty to capability uncertainty, arguing that trust cannot be static because the AI functions as a living, sociotechnical system with evolving boundaries~\cite{yang2020re}. Systematic reviews posit that trust evolves through ``contestability'', the ability of the user to challenge the outputs of the system and probe its boundaries~\cite{tun2025trust}. Similarly, recent frameworks describe trust as a mediator in a feedback loop (Input--Mediator--Output--Input), maintained through active monitoring processes rather than blind faith~\cite{mcgrath2025collaborative}.

This shift reframes the role of Interpretable AI. The ``black box'' nature of many deep learning models is a significant barrier to their adoption in healthcare, a field where accountability and justification are paramount~\cite{sivaraman2023ignore, vellido2020importance}. Consequently, interpretability in AI has become a major focus~\cite{arrieta2020explainable, holzinger2019causability}. Common techniques include feature importance methods like SHAP (SHapley Additive exPlanations) or LIME (Local Interpretable Model-agnostic Explanations), which highlight the factors driving a prediction~\cite{lundberg2017unified, ribeiro2016should}. However, studies show that standard ``feature importance'' explanations often fail to improve decision outcomes because they increase cognitive load without improving the user's ability to detect errors~\cite{wang2021explanations, nagendran2023quantifying, ghassemi2021false}. Panigutti et al. address this via a progressive disclosure interface that presents natural language summaries before raw data to mitigate information overload~\cite{panigutti2023co}. Clinicians do not necessarily need mathematical transparency. They need clinical legitimacy, interpretability that aligns with medical ontology and clinical reasoning structures~\cite{tonekaboni2019clinicians, burgess2023healthcare, calisto2025personalized}. Researchers argue for legibility and cross-checking, where AI serves as one data signal among many that clinicians cross-reference against patient history~\cite{thieme2023designing, thieme2025challenges}.

Moreover, trust construction is often an exercise in agency and alignment. Cai et al. found that pathologists need to understand the subjectivity and ``medical personality'' of the AI to treat it as a colleague rather than a tool~\cite{cai2019hello}. Goh et al. further demonstrate that clinicians are willing to modify decisions based on AI recommendations if the system engages in a transparent reasoning process rather than static output~\cite{goh2025physician}. Ideally, interactive systems should support forward reasoning, where the AI extends the user's thought process, engendering deeper engagement than those that simply provide a retrospective justification~\cite{reicherts2025ai, rajashekar2024human}. Our study builds on this by identifying specific behavioral mechanisms (such as adversarial verification) by which experts construct trust in high-stakes environments.

\subsection{Impact of Clinical Context and Expertise on AI Interaction}
A critical but under-explored dimension of AI deployment is how clinical expertise modulates interaction and reliance. The one-size-fits-all approach to AI design is increasingly scrutinized as insufficient for diverse medical teams. Calisto et al. demonstrated that the optimal communication style of an AI varies by user: junior clinicians benefit from assertive guidance that provides structure, while experts prefer suggestive support that respects their autonomy and experience~\cite{calisto2023assertiveness, calisto2025personalized}.

For junior clinicians, AI introduces the risk of de-skilling or automation bias. Gaube et al. found that novices are highly susceptible to adhering to incorrect AI advice even when they claim to be skeptical, due to a lack of confidence in their own judgment~\cite{gaube2021ai}. Klingbeil et al. further illustrate that without domain expertise, users demonstrate ``algorithm appreciation'', favoring AI advice even when it contradicts contextual information~\cite{klingbeil2024trust}. Bienefeld et al. warn that without careful design, automation can atrophy the implicit knowledge required for holistic care~\cite{bienefeld2024human, bienefeld2025ai}. However, properly designed systems can act as cognitive scaffolding. In this mode, the AI helps novices structure their learning and organizes complex data without removing the cognitive effort required for mastery~\cite{wolf2025clinical, thieme2025challenges}.

Conversely, senior clinicians often approach AI with skepticism or algorithmic resistance. They frequently view it as a tool for efficiency rather than diagnosis~\cite{bach2023if}. They engage in adversarial behaviors, utilizing the AI to challenge their own hypotheses or to double-check against errors~\cite{thieme2025challenges, goh2025physician}. While some studies suggest experts might reject AI that conflicts with their intuition, others suggest they are willing to modify decisions if the AI provides transparent, evidence-based counter-arguments that they can verify~\cite{goh2025physician, lu2024does}. Our work extends this literature by providing empirical evidence of these distinct behavioral modes within a single system evaluation. We demonstrate how AICare supports the learning of the novice via scaffolding while simultaneously withstanding the scrutiny of the expert.

\section{The AICare System}

\begin{figure*}[!ht]
  \centering
  \includegraphics[width=0.9\linewidth]{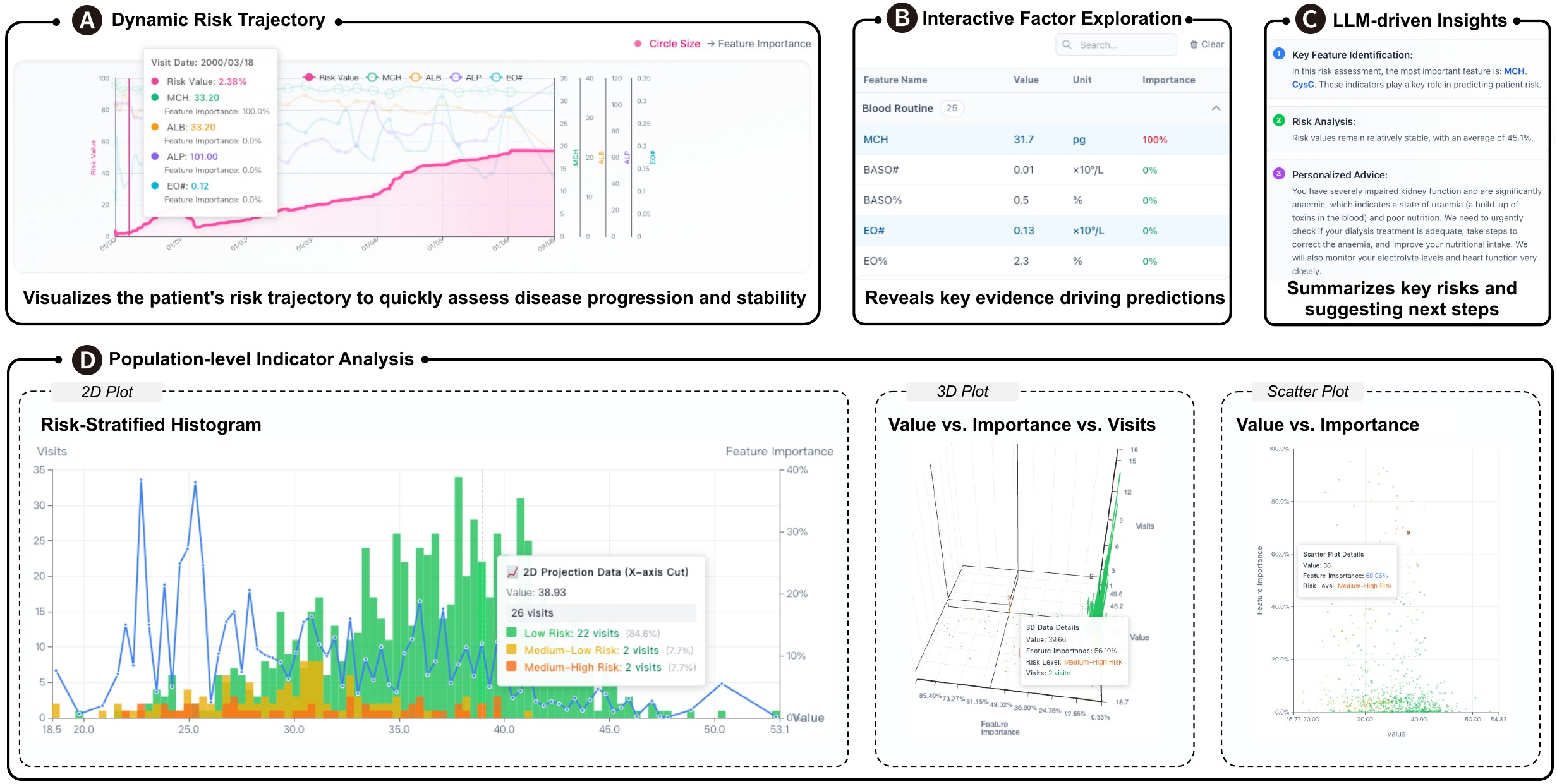}
  \caption{The AICare system interface, designed as an interactive and interpretable AI copilot. The dashboard features (A) a dynamic risk trajectory visualization over the patient's visits that externalizes memory-intensive trend analysis, (B) an interactive list of critical risk factors with their adaptive importance scores enabling immediate drill-down into evidence, (C) an LLM-driven diagnostic recommendation synthesizing key findings, and (D) a population-level indicator analysis providing cohort context for a selected feature. Interaction patterns reveal that these features support dual strategies: acting as cognitive scaffolding for novices to structure analysis, while enabling experts to perform ``adversarial verification'' of the model's logic.}
  \label{fig:aicare_platform}
\end{figure*}

AICare is a web-based, interactive, and interpretable AI copilot currently deployed and integrated into the daily clinical workflow of the participating hospitals to support clinicians in diagnostic risk assessment. Its architectural design was guided by a research-through-design approach, involving multiple iterative cycles of development and pilot deployment with senior nephrologists and obstetricians across six years.

During early stakeholder interviews, clinicians explicitly rejected models that output static risk scores, citing that such scores lacked the ``narrative context'' of disease progression. This specific clinical requirement drove our underlying model selection. We moved away from standard MLPs (multilayer perceptrons) or tree-based models, which were initially considered, toward a time-aware architecture capable of visualizing longitudinal changes, a direct response to the clinicians' need to verify the ``trajectory'' rather than just the ``state'' of a patient.

\subsection{Underlying Interpretable Predictive EHR Model}

Feedback from our pilot deployments showed that clinicians do not rely on abstract feature weight; they require clinical legitimacy where they can intuitively verify that the model's logic aligns with the patient's history. To meet this requirement for longitudinal transparency, we adapted the ConCare architecture~\cite{ma2020concare,ma2023mortality} to our AICare model as the analytical engine. This design decision was made to address two specific characteristics of real-world EHR data identified by our clinical partners:
\begin{enumerate}[leftmargin=*, label=(\arabic*), topsep=0pt]
    \item \textbf{Handling irregular sampling via multi-channel extraction.} Clinical data is sparse and irregularly sampled. Unlike standard RNNs which assume fixed time steps, our implementation uses a bidirectional Gated Recurrent Unit (bi-GRU) to capture the varying time-lapses between visits. This ensures that the model treats a gap of one week differently than a gap of one year, while simultaneously grounding these dynamics in static baseline features (e.g., demographics) processed through a separate multilayer perceptron (MLP).
    \item \textbf{Dynamic feature attribution via attention mechanisms.} Clinicians emphasized that a biomarker (e.g., Creatinine) might be irrelevant at baseline but critical during an acute flare-up. To model this, we utilized an adaptive feature importance recalibration module. By treating the patient embedding as a ``health context query'', the model uses a multi-head self-attention mechanism to assign personalized importance scores that change over time. This allows the system to highlight different risk factors at different stages of the disease progression, matching the clinician's evolving focus.
\end{enumerate}
This architecture allows AICare not only to predict a future risk (e.g., 1-year mortality) but also to generate a dynamic, personalized, and time-aware explanation for its prediction, which forms the basis for our interactive visualizations.

\subsection{System Features}
AICare's interface is organized into a primary individual patient dashboard (Figure~\ref{fig:aicare_platform}A, B, C) and a secondary population analysis view (Figure~\ref{fig:aicare_platform}D). Each component is designed to facilitate a step in the clinical reasoning process, from gaining a high-level overview to scrutinizing specific evidence and placing it in a broader context.

\paragraph{Dynamic risk trajectory visualization.}
Instead of presenting a single, static risk score, AICare visualizes the patient's predicted risk as a longitudinal trajectory over the course of their clinical visits (Figure~\ref{fig:aicare_platform}A). This time-series graph provides clinicians with an immediate, holistic view of the patient's health status, revealing trends, periods of stability, or moments of rapid deterioration. Overlaid on the main risk curve are the temporal plots of the most influential clinical indicators, such as hemoglobin or albumin. Clinicians can hover over any point on the trajectory to inspect a detailed tooltip showing the precise risk value and the corresponding values of key features for that specific visit. The size of each feature's marker on the graph visually encodes its importance, which is defined as its contribution to the final prediction, scaled as a value between 0 and 1. This visual encoding of feature weights makes the model's dynamic reasoning immediately apparent.

\paragraph{Interactive list of critical risk factors.}
This module (Figure~\ref{fig:aicare_platform}B) offers a detailed, local explanation for the prediction corresponding to the latest clinical visit. It presents a comprehensive list of all clinical features, which are grouped by category (e.g., blood routine), sorted by their predictive importance, and fully searchable for ease of access. For each feature, the interface displays its measured value, unit, and calculated importance percentage. Selecting a feature in the interactive list instantly overlays its historical trend line onto the main risk trajectory graph (Figure~\ref{fig:aicare_platform}A), allowing clinicians to visually correlate a specific indicator's fluctuations with changes in the patient's overall risk.

\paragraph{LLM-driven diagnostic recommendation.}
This component (Figure~\ref{fig:aicare_platform}C) synthesizes the quantitative outputs of the model into a concise, narrative summary designed for clinical decision-making. The LLM-generated text is structured into three parts: ``Key Feature Identification'', which highlights the most critical indicators driving the risk assessment; ``Risk Analysis'', which offers a qualitative summary of the risk trajectory; and ``Personalized Advice'', which translates the findings into a preliminary clinical interpretation and suggests potential areas for further investigation. This feature bridges the gap between the model's data-driven findings and actionable clinical thought processes.

\paragraph{Population-level indicator analysis.}
To contextualize an individual's data, this view (Figure~\ref{fig:aicare_platform}D) enables clinicians to compare a specific indicator against cohort-level trends derived by performing inference on a random subset of 100 samples and aggregating results across all patient visits. This process establishes a global baseline that visualizes the multi-dimensional relationships between three key variables: the feature's measured value, its calculated importance, and the predicted risk probability. These relationships are rendered in interactive formats, including 2D, 3D, and scatter plots, to reveal how specific feature magnitudes correlate with overall risk. By positioning the current patient within this distribution, clinicians can ascertain whether a flagged feature represents a consistent population-level risk pattern or an atypical anomaly, thereby strengthening the evidentiary basis of their decisions.

\subsection{Implementation Details}

\paragraph{Real-world clinical datasets and predictive tasks.}

To ensure the system's robustness and clinical validity, AICare was implemented and evaluated using three distinct real-world Electronic Health Record (EHR) datasets sourced from three different departments. These datasets cover two high-stakes specialties and are characterized by sparse, high-dimensional, and irregularly sampled time-series data (see Table~\ref{tab:dataset_stats} for statistics). Detailed inclusion/exclusion criteria and comprehensive feature statistics for all cohorts are provided in Appendix~\ref{app:cohort_selection} and Appendix~\ref{app:dataset_features_stats}.

\begin{itemize}[leftmargin=*, topsep=0pt]
    \item \textbf{Nephrology (ESRD Mortality Risk):} We utilized datasets from XY Hospital (a large regional general hospital) and BS Hospital (a top-tier national hospital). These collections track patients with End-Stage Renal Disease (ESRD) undergoing long-term dialysis. The data characteristics include static demographics (e.g., age, gender, primary disease) and longitudinal dynamic laboratory values (e.g., albumin, creatinine, potassium) recorded during varying follow-up intervals. The specific task is to predict the rolling risk of one-year all-cause mortality at each patient visit.
    \item \textbf{Obstetrics (Preterm Birth Risk):} We utilized a dataset from BC Hospital (a top-tier national hospital). This dataset comprises records of women with multiple gestation pregnancies. The data characterizes the prenatal journey through maternal demographics and time-varying clinical indicators (e.g., cervical length, uterine contractions, laboratory results). The predictive task is to assess the risk of spontaneous preterm birth (delivery $<37$ weeks) at each prenatal check-up.
\end{itemize}
\begin{table}[!ht]
\centering
\caption{Descriptive statistics of the three EHR datasets used to train the predictive models and source the patient cases for the study.}
\label{tab:dataset_stats}
\resizebox{\columnwidth}{!}{%
    \begin{tabular}{@{}lrrr@{}}
    \toprule
    \textbf{Statistic} & \textbf{XY-Nephrology} & \textbf{BS-Nephrology} & \textbf{BC-Obstetrics} \\
    \midrule
    Total Patients & 1,404 & 341 & 5,315 \\
    Eligible Patients & 1,224 & 280 & 3,577 \\
    Total Records Loaded & 7,461 & 8,938 & 43,038 \\
    Avg. Visits (Timesteps) & 6.46 & 19.20 & 12.03 \\
    Positive Outcome Ratio & 26.80\% & 44.29\% & 54.12\% \\
    \midrule
    \textbf{Feature Counts} & & & \\
    \quad Static & 7 & 23 & 29 \\
    \quad Dynamic & 33 & 66 & 50 \\
    \midrule
    \textbf{Selected Cases for Study} & & & \\
    \quad Total Selected & 10 & 10 & 10 \\
    \quad Positive Outcomes & 5 & 3 & 5 \\
    \quad Negative Outcomes & 5 & 7 & 5 \\
    \bottomrule
    \end{tabular}
}
\end{table}

\begin{figure*}[!ht]
  \centering
  \includegraphics[width=0.9\linewidth]{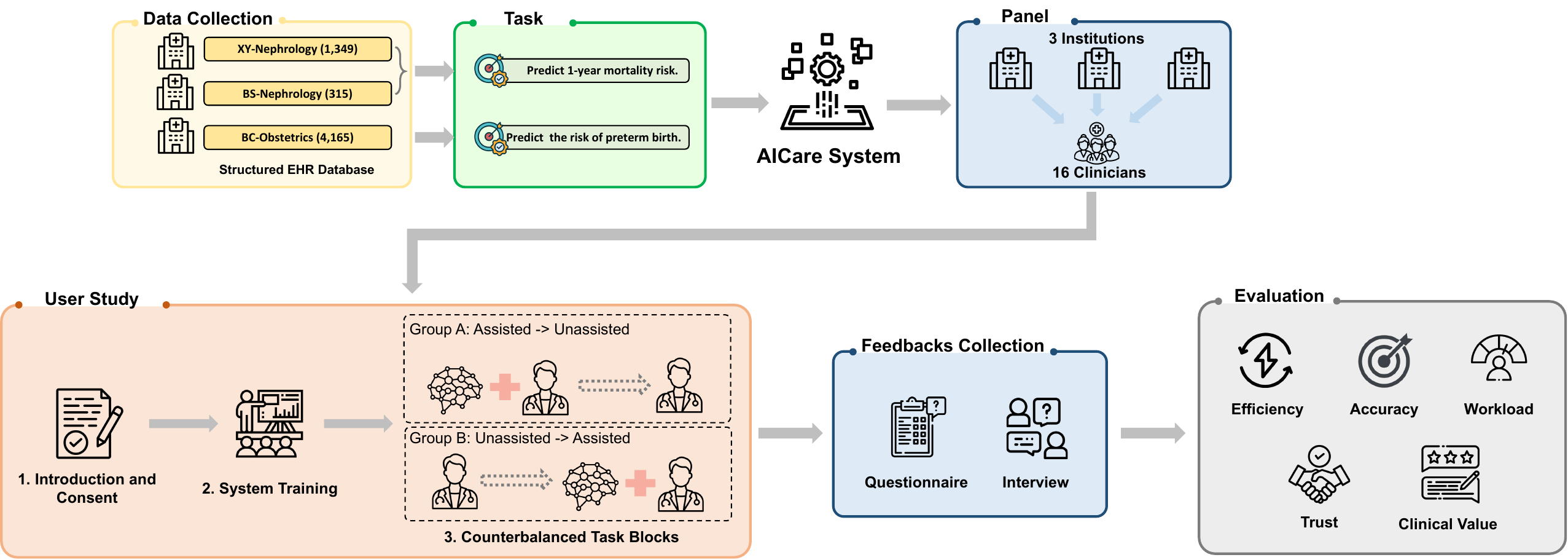}
  \caption{An overview of the user study methodology, from data collection to evaluation. We utilized structured EHR data from three institutions across two specialties (Nephrology and Obstetrics) to train the AICare system for two predictive tasks. The study involved 16 clinicians in a rigorous within-subjects, counterbalanced design. Participants completed risk assessment tasks under two conditions: AI-assisted (using AICare) and unassisted (a baseline). We collected quantitative and qualitative data through post-task questionnaires and semi-structured interviews to evaluate AICare's impact on efficiency, accuracy, cognitive workload, trust, and perceived clinical value. Quantitative results demonstrate that AICare significantly reduces cognitive workload by offloading the mental burden of synthesizing longitudinal data, without compromising diagnostic accuracy.}
  \label{fig:user_study}
\end{figure*}

For each specialty (Nephrology and Obstetrics), we curated a set of 10 real, anonymized patient cases for the study. These cases were drawn from a held-out test set of their respective datasets to ensure the model had not been trained on them. We employed a stratified random sampling strategy based on ground-truth labels, thus including a mix of both positive and negative outcomes (e.g., mortality/survival, preterm/term birth). Furthermore, the selection included cases where the model's prediction was correct as well as cases where it was incorrect. This was done intentionally to observe how clinicians would interact with the system, calibrate their trust, and negotiate potential disagreements with the AI's recommendations. The ground truth outcome for each case was known to the researchers but not revealed to the participants during the study.

\paragraph{Validation of predictive performance benchmarks.}
To ensure that study participants interacted with a clinically competent system, we rigorously evaluated the underlying predictive models prior to deployment. Using a stratified 10-fold cross-validation strategy, the Nephrology model (predicting 1-year mortality) achieved an Area Under the Receiver Operating Characteristic (AUROC) of 0.898 and an Area Under the Precision-Recall Curve (AUPRC) of 0.906, surpassing recent state-of-the-art benchmarks, such as the Transformer-based approach by Xu et al. (AUROC=0.877)~\cite{xu2024machine}. Similarly, the Obstetrics model (predicting preterm birth) achieved an AUROC of 0.740 and AUPRC of 0.775, surpassing specialized baselines like Bacsaran et al. (AUROC=0.71)~\cite{bacsaran2025role}. Establishing this high level of predictive fidelity was a prerequisite for the user study, ensuring that the risk trajectories and feature attributions presented to clinicians in AICare were derived from a robust and reliable analytical engine.

\paragraph{Training details for EHR models.}
Following the PyEHR toolkit~\cite{zhu2024pyehr}, the modeling pipeline transforms these raw, longitudinal EHR datasets into calibrated, predictive models. The process includes feature engineering, handling of missing values using methods such as Last Observation Carried Forward (LOCF)~\cite{woolley2009last} and median imputation, feature normalization, and balancing of class distributions through oversampling. Complete details regarding data preprocessing, model architecture, and hyperparameter configuration are provided in Appendix~\ref{app:ehr_preprocessing_training}.

The underlying EHR model for interpretability was implemented in PyTorch and PyTorch Lightning and was trained using the Adam optimizer, with an early stopping strategy based on the Area Under the Precision-Recall Curve (AUPRC) on a validation set to prevent overfitting. We employ a post-hoc optimization stage, which involves two key processes: first, we apply Temperature Scaling to calibrate the model's raw outputs, ensuring the predicted probabilities are more reliable and intuitive for clinical interpretation~\cite{guo2017calibration}. Second, we determine an optimal decision threshold by maximizing the F-beta score on the validation set, allowing us to tune the model's sensitivity and specificity to align with the specific diagnostic needs of each clinical scenario.

\paragraph{Development of AICare system.}
The AICare system is engineered as a decoupled web application. The backend is developed using FastAPI~\cite{tiangolo2018fastapi}, a modern Python web framework chosen for its high performance and native support for asynchronous operations, which is crucial for handling model inference requests without blocking the user interface. Data persistence is managed by a SQLite database via the SQLAlchemy ORM. The backend loads pretrained AICare models and exposes a set of RESTful API endpoints that serve as the analytical core of the system. The frontend is a single-page application built with Vue 3~\cite{vue3} and the ElementPlus UI library, ensuring a modular and responsive design. Interactive visualizations are rendered using Apache ECharts. For the LLM-driven diagnostic recommendation, the frontend directly queries the DeepSeek API using its official client, feeding it a structured prompt (detailed in Appendix~\ref{app:llm_generate_advice_prompt}) containing the model's key findings to generate a concise narrative summary. To mitigate the risk of hallucinations, we employed a constraint-based prompting strategy that explicitly instructs the LLM to ground its narrative strictly in the provided quantitative evidence. Given that inconsistencies may still arise, the system relies on a human-in-the-loop validation model, allowing our study to empirically examine how clinicians perceive and scrutinize these AI-generated insights.

\paragraph{Hardware and software configuration.}
All EHR model training experiments were run on an Apple Mac Studio M3 Ultra with 512GB of RAM. The deployment of the AICare system was also based on the Mac Studio. The primary software stack comprises Python 3.12, PyTorch 2.6.0, PyTorch Lightning 2.5.1, and Transformers 4.50.0. Experiments were conducted between August 1, 2025, and September 1, 2025. The LLM-generated recommendations utilize the DeepSeek-V3~\cite{liu2024deepseekv3}, DeepSeek-R1~\cite{guo2025deepseekr1}, DeepSeek-V3.1~\cite{deepseekv31} model via its official API. Interview recordings were transcribed using Gemini 2.5 Pro~\cite{comanici2025gemini}. To strictly control for hallucinations or errors, the generated text was cross-validated by two researchers prior to inclusion in the qualitative analysis.

\section{User Study Methodology}

\begin{table*}[!ht]
\centering
\caption{Participant demographics and background ($N=16$). Participants were from three hospital sites (XY, BC, BS). Condition: ``AI-first'' participants used the AI system before their own analysis; ``Human-first'' participants did the reverse. ``AI Fam.'' (AI Familiarity): Self-reported on a 5-point Likert scale (1: Completely Unfamiliar, 5: Expert). Exp. (Experience).}
\label{tab:participants}
\resizebox{0.95\linewidth}{!}{%
\begin{tabular}{@{}llccclcccc@{}}
\toprule
\textbf{ID} & \textbf{Role} & \textbf{Exp. (Yrs)} & \textbf{Age Group} & \textbf{Gender} & \textbf{Department} & \textbf{Condition} & \textbf{Interviewed} & \textbf{AI Fam.} & \textbf{Prior AI Use} \\
\midrule
XY-01A & Assoc. Chief Physician & 16-20 & 40-49 & F & XY-Nephrology & AI-first & No & 2 & Occasionally \\
XY-02B & Nurse & 11-15 & 30-39 & F & XY-Nephrology & Human-first & Yes & 2 & Occasionally \\
XY-03A & Attending Physician & 6-10 & 30-39 & F & XY-Nephrology & AI-first & Yes & 3 & Never \\
XY-04B & Assoc. Chief Physician & 16-20 & 40-49 & M & XY-Nephrology & Human-first & Yes & 2 & Never \\
XY-05A & Chief Physician & 16-20 & 40-49 & M & XY-Nephrology & AI-first & No & 2 & Never \\
XY-06B & Assoc. Chief Nurse & 11-15 & 30-39 & F & XY-Nephrology & Human-first & Yes & 2 & Occasionally \\
XY-07A & Resident Physician & 1-5 & 30-39 & M & XY-Nephrology & AI-first & Yes & 3 & Frequently \\
XY-08B & Chief Physician & >20 & >50 & F & XY-Nephrology & Human-first & Yes & 3 & Never \\
\midrule
BS-01A & Medical Student & 1-5 & <30 & F & BS-Nephrology & AI-first & No & 3 & Frequently \\
BS-02B & Medical Student & 1-5 & <30 & F & BS-Nephrology & AI-first & No & 3 & Frequently \\
BS-03A & Resident Physician & 1-5 & <30 & F & BS-Nephrology & Human-first & No & 3 & Daily \\
BS-04B & Chief Physician & >20 & 40-49 & F & BS-Nephrology & Human-first & Yes & 5 & Never \\
BS-05A & Intern & 1-5 & <30 & F & BS-Nephrology & AI-first & No & 5 & Occasionally \\
\midrule
BC-01A & Attending Physician & 6-10 & 30-39 & F & BC-Obstetrics & AI-first & Yes & 3 & Frequently \\
BC-02B & Assoc. Chief Physician & 11-15 & 30-39 & F & BC-Obstetrics & Human-first & No & 4 & Occasionally \\
BC-03A & Attending Physician & 6-10 & 30-39 & F & BC-Obstetrics & AI-first & Yes & 3 & Frequently \\
\bottomrule
\end{tabular}%
}
\end{table*}

To investigate our research questions, we designed and conducted a within-subjects user study with a counterbalanced order of conditions (see Figure~\ref{fig:user_study}). This approach allowed us to rigorously evaluate the impact of AICare on clinicians' performance and perceptions while minimizing inter-participant variability and gathering rich qualitative insights into their experience. The study protocol received ethical approval from the institutions' research ethics committees.

\subsection{Participants}
We recruited a total of 20 medical professionals through professional contacts at three hospital sites representing two different tiers of the healthcare system. The participants were divided into two cohorts based on the study phase: a primary cohort for the main user study ($N=16$) and a supplementary cohort for in-depth interviews ($N=4$). All participants were compensated for their time. The study protocol received ethical approval from the Institutional Review Board (IRB) at each participating hospital (see Section~\ref{sec:ethics}).

\paragraph{Primary user study cohort ($N=16$).}
Participants details for the main within-subjects experiment are summarized in Table~\ref{tab:participants}. This group included:
\begin{itemize}[leftmargin=*, topsep=0pt]
    \item \textbf{XY Hospital (Nephrology):} Department of Nephrology, Affiliated Xuzhou Municipal Hospital of Xuzhou Medical University. A large regional hospital, from which we recruited 8 professionals, including physicians, nurses, and residents.
    \item \textbf{BS Hospital (Nephrology):} Department of Nephrology, Peking University Third Hospital. A top-tier national hospital, from which we recruited 5 professionals, including physicians, residents, and medical students.
    \item \textbf{BC Hospital (Obstetrics):} Department of Obstetrics and Gynecology, Peking University Third Hospital. A top-tier national hospital specializing in women and children's health, from which we recruited 3 attending or associate chief physicians.
\end{itemize}

Participants' clinical experience ranged from 1 to over 20 years ($M$ around 11 with $SD$ around 7). Their self-reported familiarity with AI-based CDSS varied ($M$ around 2.8, $SD$ around 1.0 on a 5-point scale), with senior clinicians often reporting less prior use than junior clinicians.

\paragraph{Supplementary interview cohort ($N=4$).}

To complement the quantitative metrics with granular insights into clinical workflow integration, we recruited an additional group of 4 participants specifically for semi-structured interviews. These participants did not perform the timed diagnostic tasks but provided detailed narrative feedback on system usability and cognitive processing. Detailed demographics for this group are presented in Table~\ref{tab:more_participants}.

\begin{table*}[!ht]
\centering
\caption{Demographics of the additional participant group (N=4) interviewed for the discussion analysis. ``Condition'' indicates the counterbalanced order: ``AI-first'' participants used the AICare system prior to the conventional tabular interface, while ``Table-first'' participants followed the reverse order. ``AI Fam.'': Self-reported AI familiarity on a 1--5 scale.}
\label{tab:more_participants}
\resizebox{0.9\linewidth}{!}{%
\begin{tabular}{@{}llccclcccc@{}}
\toprule
\textbf{ID} & \textbf{Role} & \textbf{Exp. (Yrs)} & \textbf{Age Group} & \textbf{Gender} & \textbf{Department} & \textbf{Condition} & \textbf{Interviewed} & \textbf{AI Fam.} & \textbf{Prior AI Use} \\
\midrule
BC-A1 & Resident Clinician & 1-5 & <30 & F & BC-Obstetrics & Table-first & Yes & 3 & Never \\
BC-A2 & Attending Clinician & 1-5 & 30-39 & M & BC-Obstetrics & AI-first & Yes & 5 & Daily \\
BC-A3 & Attending Clinician & 6-10 & 30-39 & M & BC-Obstetrics & Table-first & Yes & 4 & Frequently \\
BC-A4 & Medical Student & <1 & <30 & F & BC-Obstetrics & AI-first & Yes & 4 & Never \\
\bottomrule
\end{tabular}%
}
\end{table*}

\subsection{Study Scenarios and Conditions}
To evaluate generalizability across distinct medical contexts, we designed two study scenarios corresponding to the specific clinical tasks: (1) Nephrology, focusing on the 1-year mortality risk for patients with End-Stage Renal Disease (ESRD), and (2) Obstetrics, focusing on the risk of spontaneous preterm birth during prenatal care.

Participants analyzed 10 anonymized patient cases specific to their specialty. The study employed a within-subjects design with two experimental conditions to counterbalance the order of conditions and case sets to control for individual variability and order effects:
\begin{itemize}[leftmargin=*, topsep=0pt]
    \item \textbf{Unassisted (Control):} Participants reviewed a static, non-interactive digital summary of the patient's EHR data. This interface presented the same core information (demographics, visit history, lab results) available in AICare but without any predictive analytics or interactive visualizations.
    \item \textbf{AI-assisted (AICare):} Participants used the full AICare interface, which was populated with the same underlying patient data and our trained predictive models.
\end{itemize}

\subsection{Procedure}
The study session for each participant lasted approximately 60 to 75 minutes and was conducted one-on-one with a researcher in a quiet office. The procedure followed a structured protocol:
\begin{enumerate}[leftmargin=*, label=(\arabic*), topsep=0pt]
    \item \textbf{Introduction and Consent (approx. 5 min).} The researcher explained the study's purpose, obtained written informed consent, and the participant completed a background questionnaire (Appendix~\ref{app:participant_background}).
    \item \textbf{System Training (approx. 10 min).} Participants received a standardized tutorial (following the script in Appendix~\ref{app:guidance_script}) on the AICare interface using a sample case not included in the main study tasks. This ensured a consistent level of proficiency with the system's features. During this session, we explicitly articulated the probabilistic nature of the AI and the possibility of errors, directing participants' attention to the interface's disclaimer that all AI outputs are provided as ``preliminary clinical interpretations'' requiring human verification.
    \item \textbf{Counterbalanced Task Blocks (approx. 40 min).} Participants were randomly assigned to one of the following two groups (AI-first or Human-first):
        \begin{itemize}[leftmargin=*, topsep=0pt]
            \item \textit{Group A (n=9, AI-first):} Analyzed 5 cases using AICare, then 5 cases in the unassisted condition.
            \item \textit{Group B (n=7, Human-first):} Completed the tasks in the reverse order.
        \end{itemize}
        For each case, the participant's task was to review the patient's record and provide a risk assessment (e.g., ``What is the likelihood of mortality in the next year?'') on a 4-point Likert scale (Low, Low-Medium, Medium-High, High), along with their confidence in that assessment on a 5-point Likert scale (see Appendix~\ref{app:case_questionnaire}). We logged the time taken for each case.
    \item \textbf{Workload Assessment.} Immediately after completing each block of 5 cases, participants filled out the NASA-Task Load Index (NASA-TLX) questionnaire to assess their perceived cognitive workload for that condition~\cite{hart2006nasa}.
    \item \textbf{Post-Task Questionnaires (approx. 10 min).} After completing all 10 cases, participants filled out the System Usability Scale (SUS)~\cite{lewis2018sus} and the Trust in Automation scale for the AICare system. In addition, participants completed a ``AICare Feature Feedback'' form (Appendix~\ref{app:feature_feedback}), where they rated the specific usefulness of AICare's individual modules.
    \item \textbf{Semi-Structured Interview (approx. 10-15 min).} The session concluded with an audio-recorded interview with participants who consented (n=9). The interview, detailed in Appendix~\ref{app:interview}, explored their overall impressions of AICare, focusing on trust, usability, workflow integration, and the system's clinical value.
\end{enumerate}

\subsection{Measures and Data Analysis}

Our study employs a combination of quantitative and qualitative analyses to provide a comprehensive evaluation. We quantitatively analyze performance metrics and survey responses, and qualitatively analyze data from semi-structured interviews.

\paragraph{Quantitative measures.}
We collected data across five key dimensions: efficiency, accuracy, workload, usability, and trust. Each measure is detailed below.

\begin{itemize}[leftmargin=*, topsep=0pt]
    \item \textbf{Efficiency.} We measure efficiency as the task completion time in seconds for each case assessment. The timer starts when a participant is first presented with a case and stops upon completing a valid diagnosis and submitting their confidence rating. Time spent in discussions with the research team for procedural clarifications or to provide feedback on the user interface is excluded from this measurement. This ensures that the data reflects only the time dedicated to the assessment task itself.

    \item \textbf{Accuracy.} Participants assessed patient risk using a 4-point Likert scale (A: 0--25\%, B: 25--50\%, C: 50--75\%, D: 75--100\%). To compute accuracy, we first binarize these assessments into a low-risk category (A and B) and a high-risk category (C and D). This binarized assessment is then compared against the patient's actual binary outcome (ground truth) to determine correctness. We calculate the accuracy for each participant as the percentage of correct assessments under each condition (unassisted and AI-assisted).

    \item \textbf{Workload.} We assess cognitive workload using the NASA Task Load Index (NASA-TLX)~\cite{hart1988development,hart2006nasa} (see Appendix~\ref{app:nasa_tlx}). Participants rated their perceived workload across six dimensions (mental demand, physical demand, temporal demand, performance, effort, and frustration) on a continuous scale (e.g., ``How much mental and perceptual activity was required?'' for mental demand, and ``How insecure, discouraged... did you feel?'' for frustration). In our analysis, we use the raw TLX score, which is the unweighted average of the six ratings, resulting in a composite workload score from 0 to 100 for each condition.

    \item \textbf{Usability.} We evaluate the usability of the AICare system with the System Usability Scale (SUS)~\cite{brooke1996sus} (Appendix~\ref{app:sus}), a standardized 10-item questionnaire with a 5-point Likert scale ranging from 1 (``Strongly disagree'') to 5 (``Strongly agree''). Items include statements such as ``I found the system unnecessarily complex'' and ``I thought the system was easy to use''. The final SUS score is calculated by converting responses according to standard procedure: for positively worded items (1, 3, 5, 7, 9), we subtract 1 from the score; for negatively worded items (2, 4, 6, 8, 10), we subtract the score from 5. The sum of these converted scores is then multiplied by 2.5, yielding a final usability score from 0 to 100~\cite{bangor2008empirical}.

    \item \textbf{Trust.} We measure participants' trust in the AI system using the Trust in Automation Scale~\cite{jian2000foundations} (Appendix~\ref{app:trust}). This scale consists of 12 items rated on a 7-point Likert scale (1=``Strongly disagree'', 7=``Strongly agree''), including questions such as ``The system is deceptive'' and ``I am confident in the system.'' Several items (1, 4, 6, 9, 11) are negatively worded and are reverse-scored by subtracting their value from 8. The final trust score is the average of all 12 item scores after reversal, resulting in a value between 1 and 7.

    \item \textbf{Interaction Behaviors.} To verify the depth of cognitive engagement, we analyzed screen recordings to extract granular interaction metrics. We quantified specific actions as proxies for verification intensity: (1) \textit{List paging:} Frequency of scrolling or paging through the risk factor list, serving as a proxy for information seeking beyond the top-ranked features; (2) \textit{Curve hovering:} Frequency of hovering over specific time points on the risk trajectory. In time-series visual analytics, hovering is often correlated with detailed value inspection and local trend verification~\cite{heer2012interactive}; (3) \textit{Cross-view comparison:} Frequency of moving the cursor between the historical trend view and current status view, indicating an active process of synthesizing longitudinal context with cross-sectional evidence.
\end{itemize}

\paragraph{Qualitative measures.}
The nine semi-structured interviews (guiding questions provided in Appendix~\ref{app:interview}) were transcribed and analyzed using a reflexive thematic analysis approach~\cite{braun2006using}. To ensure accuracy, the AI-generated transcripts were manually cross-validated by two researchers against the original audio recordings prior to analysis. The two researchers independently coded the transcripts, identified initial themes, and then collaboratively met to discuss, merge, and refine these themes into a final thematic structure that captured the core aspects of the clinicians' experience with AICare.

\paragraph{Statistical analysis.}
We analyzed the collected data using statistical methods appropriate for the experimental design. For the within-subjects factors (efficiency, accuracy, and workload), we conducted a repeated measures analysis of variance (RM-ANOVA). In cases where the data violated the assumption of normality, we employed the Aligned Rank Transform (ART) for non-parametric factorial analysis~\cite{wobbrock2011aligned}, a robust method for examining main effects and interactions. The assumption of sphericity was assessed using Mauchly's test, and when violated, Greenhouse–Geisser (GG) corrections were applied to adjust the degrees of freedom; all reported $p$-values for within-subject effects reflect the GG-adjusted tests when applicable. For the between-subjects factor of task order (AI-assisted first vs. unassisted first), we used an independent samples t-test to examine its effect on measures such as trust. For all analyses, we report effect sizes (partial eta-squared, $\eta_p^2$, for ANOVA; Cohen's d for t-tests). Post-hoc pairwise comparisons were adjusted using Holm's correction for multiple comparisons. All statistical analyses were performed in R, interfaced through the \texttt{rpy2} library in Python, utilizing the \texttt{ARTool}~\cite{wobbrock2011aligned} and \texttt{afex} packages.

\subsection{Ethical Considerations}
\label{sec:ethics}
Our research was conducted in full compliance with the ACM Policy on Research Involving Human Participants and Subjects and the ethical principles of the Declaration of Helsinki. All study procedures, including participant recruitment and informed consent, were reviewed and approved by the Medical Science Research Ethics Committees of the participating hospitals prior to study commencement.

We obtained written informed consent from all clinician participants. The consent process ensured they understood the study's purpose, the voluntary nature of their participation, and their right to withdraw at any time. All collected data, including audio recordings and survey responses, were anonymized to protect participant confidentiality. Participants were compensated for their professional time.

\section{Results Analysis}
\begin{figure*}[!ht]
  \centering
  \includegraphics[width=0.9\linewidth]{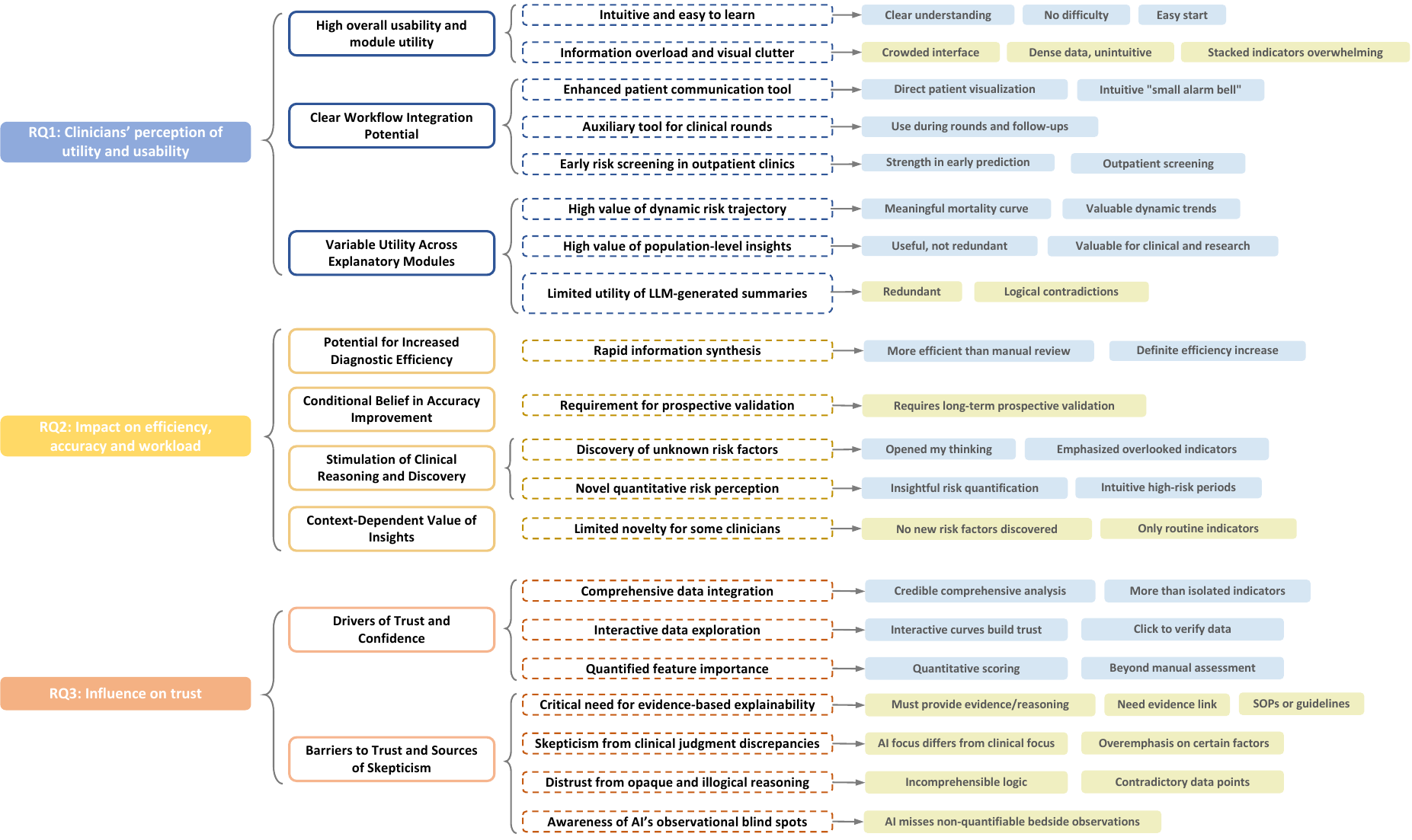}
  \caption{Thematic map of clinician interview findings. The hierarchical structure organizes key themes and sub-themes derived from semi-structured interviews. Findings are grouped by each of the three core research questions (RQs) and are supported by representative quotes from participants. Qualitative analysis confirms that transparency amplifies trust when the reasoning is clinically plausible, but accelerates rejection when the AI contradicts established medical logic.}
  \label{fig:tree}
\end{figure*}

We present our findings organized by our research questions, integrating quantitative and qualitative results to provide a comprehensive picture. The qualitative insights derived from our semi-structured interviews were synthesized using reflexive thematic analysis to capture the nuances of clinician interaction. Figure~\ref{fig:tree} presents the resulting thematic map, organizing findings into three high-level categories aligned with our research questions: (1) the perception of utility and usability, highlighting the balance between high functional value and potential visual clutter; (2) the impact on efficiency, accuracy and workflow, emphasizing the system's role in stimulating clinical reasoning versus confirming known risks; and (3) the influence on trust, distinguishing between drivers of confidence (such as interactive verification) and sources of skepticism (such as logical contradictions or blind spots).

\subsection{Analysis to RQ1}

\paragraph{AICare is perceived as useful but visually demanding, with variations across hospital tiers.}
Overall, participants rated AICare's features as highly useful for clinical practice, with a mean usefulness score of 79.69 ($SD=14.70$, Table~\ref{tab:system_feature_usefulness}). The dynamic risk trajectory visualization ($M=82.81$) and interactive list of critical risk factors ($M=81.25$) were deemed the most valuable components. Clinicians appreciated the system's ability to narrate a patient's health history. For instance, a nurse from the regional hospital (XY-02B) noted, ``Seeing the risk trend over time is far more valuable than a single score.'' Similarly, a resident from the national hospital (BS-03A) found the trend charts ``easy to understand'', noting that observing a drop in hemoglobin helped quickly associate it with clinical events like bleeding. Beyond individual diagnosis, participants envisioned specific integration points for AICare within the broader workflow: XY-02B described it as ``a small alarm bell'' to enhance patient communication during follow-ups, while XY-08B viewed it as an auxiliary tool for clinical rounds.

\begin{table}[!ht]
    \centering
    \caption{Descriptive statistics of perceived usefulness scores for system features. F1: Dynamic Risk Trajectory Visualization, F2: Interactive List of Critical Risk Factors, F3: LLM-driven Diagnostic Recommendation, F4: Population-level Indicator Analysis. Scores were rated on a scale from 0 (not at all useful) to 100 (extremely useful).}
    \label{tab:system_feature_usefulness}
\resizebox{\columnwidth}{!}{%
    \begin{tabular}{@{} l l r c r r r @{}}
        \toprule
        \textbf{Category} & \textbf{Feature} & \textbf{N} & \textbf{Mean\stddev{SD}} & \textbf{Median} & \textbf{Min} & \textbf{Max} \\
        \midrule
        Overall Score & Usefulness Score & 16 & 79.69\stddev{14.70} & 81.25 & 50.00 & 100.00 \\
        \midrule
        Individual Feature & F1 & 16 & 82.81\stddev{11.97} & 75.00 & 75.00 & 100.00 \\
        & F2 & 16 & 81.25\stddev{19.36} & 75.00 & 50.00 & 100.00 \\
        & F3 & 16 & 78.12\stddev{20.16} & 75.00 & 25.00 & 100.00 \\
        & F4 & 16 & 76.56\stddev{24.95} & 75.00 & 25.00 & 100.00 \\
        \midrule
        By Department & XY-Nephrology & 8 & 85.16\stddev{12.91} & 81.25 & 68.75 & 100.00 \\
        & BS-Nephrology & 5 & 78.75\stddev{14.39} & 81.25 & 62.50 & 100.00 \\
        & BC-Obstetrics & 3 & 66.67\stddev{15.73} & 68.75 & 50.00 & 81.25 \\
        \midrule
        By Seniority & Senior clinicians & 9 & 75.69\stddev{13.42} & 75.00 & 50.00 & 100.00 \\
        & Junior clinicians & 7 & 84.82\stddev{15.67} & 81.25 & 62.50 & 100.00 \\
        \bottomrule
    \end{tabular}
}
\end{table}

However, System Usability Scale (SUS) score revealed a noticeable divide (Table~\ref{tab:sus_scores}). While the overall score was 63.91 (marginally acceptable), clinicians at the regional hospital (XY) rated it higher ($M=72.19$) than those at national centers (BS: $M=56.00$; BC: $M=55.00$). This performance gap implies that the system's utility may be stratified by clinical expertise, institutional workflow, case complexity, or the specific model's utility. In qualitative feedback, while clinicians found the system relatively easy to learn (Q7 $M=80.00$), many found the interface unnecessarily complex (Q2 $M=45.31$, reversed) and cumbersome (Q8 $M=35.94$, reversed). For instance, several participants (XY-07A, BS-03A, BS-04B, BC-01A) criticized the dynamic risk plot when multiple indicators were overlaid, describing the chart as ``messy'', ``too narrow'' or ``overwhelming'', leading one senior nurse (XY-02B) to suggest that the core curve should be amplified to avoid information overload. Notably, while interaction patterns were consistent across domains, the visualization required domain-specific adaptation, such as anchoring the timeline to gestational weeks in obstetrics versus calendar dates in nephrology.

\begin{table}[!ht]
    \centering
    \caption{Descriptive statistics of the System Usability Scale (SUS) scores. Both the total score and individual item scores are presented on a 0-100 scale. Item labels Q1-Q10 correspond to the individual questions of the SUS, with (R) indicating a reverse-scored item.}
    \label{tab:sus_scores}
\resizebox{\columnwidth}{!}{%
    \begin{tabular}{@{} l l r c r r r @{}}
        \toprule
        \textbf{Category} & \textbf{Item} & \textbf{N} & \textbf{Mean\stddev{SD}} & \textbf{Median} & \textbf{Min} & \textbf{Max} \\
        \midrule
        Overall Score & SUS Score & 16 & 63.91\stddev{13.72} & 67.50 & 45.00 & 90.00 \\
        \midrule
        Individual Item & Q1: Frequent Use & 16 & 71.88\stddev{17.97} & 75.00 & 25.00 & 100.00 \\
        & Q2: Unnecessarily Complex (R) & 16 & 45.31\stddev{20.85} & 50.00 & 25.00 & 100.00 \\
        & Q3: Easy to Use & 16 & 73.44\stddev{21.35} & 75.00 & 25.00 & 100.00 \\
        & Q4: Need Technical Support (R) & 16 & 51.56\stddev{29.54} & 37.50 & 25.00 & 100.00 \\
        & Q5: Functions Well Integrated & 16 & 71.88\stddev{15.48} & 75.00 & 50.00 & 100.00 \\
        & Q6: Too Much Inconsistency (R) & 16 & 45.31\stddev{27.72} & 50.00 & 0.00 & 100.00 \\
        & Q7: Quick to Learn & 15 & 80.00\stddev{19.36} & 75.00 & 50.00 & 100.00 \\
        & Q8: Awkward/Cumbersome (R) & 16 & 35.94\stddev{25.77} & 25.00 & 0.00 & 100.00 \\
        & Q9: Felt Confident & 16 & 60.94\stddev{20.35} & 50.00 & 25.00 & 100.00 \\
        & Q10: Needed to Learn a Lot (R) & 16 & 35.94\stddev{27.34} & 25.00 & 0.00 & 100.00 \\
        \midrule
        By Department & XY-Nephrology & 8 & 72.19\stddev{12.57} & 68.75 & 50.00 & 90.00 \\
        & BS-Nephrology & 5 & 56.00\stddev{10.69} & 50.00 & 45.00 & 67.50 \\
        & BC-Obstetrics & 3 & 55.00\stddev{9.01} & 52.50 & 47.50 & 65.00 \\
        \midrule
        By Seniority & Senior clinicians & 9 & 67.22\stddev{15.43} & 67.50 & 47.50 & 90.00 \\
        & Junior clinicians & 7 & 59.64\stddev{10.75} & 67.50 & 45.00 & 70.00 \\
        \bottomrule
    \end{tabular}
}
\end{table}

\paragraph{The LLM-driven insights provide efficiency but lack domain specificity.}
The LLM-driven diagnostic recommendation (Figure~\ref{fig:aicare_platform}C) received mixed feedback ($M=78.12$). It was praised by some as a ``quick consult note'' (BC-01A) that streamlined the initial review. However, domain-specific limitations were evident. In obstetrics, where protocols are strictly defined by gestational age, BC-03A found the text recommendations ``not well-integrated with clinical practice'', preferring explicit risk percentages at specific timepoints (e.g., 28 or 34 weeks) over generic text. Furthermore, BS-04B described the suggestions as ``boilerplate'' and ``not very professional'', indicating that for the LLM to be truly effective in specialized centers, it must move beyond general summaries to offer guideline-specific, actionable interventions.

\paragraph{Contextual modules serve research needs over daily clinical routine.}
The population-level indicator analysis (Figure~\ref{fig:aicare_platform}D) was identified as having a distinct ``research value'' (BC-03A), scoring moderately high in perceived usefulness ($M=76.56$). Residents and clinicians (XY-07A, XY-03A) praised its ability to contextualize individual values, establish local safety thresholds (e.g., defining green/red zones), and generate hypotheses for scientific inquiry. However, the practical application of this module in a fast-paced setting was questioned; obstetrician BC-01A noted it had a ``high learning curve'' and would likely be used for ``model validation by engineers'' or retrospective research rather than point-of-care decisions. This delineates a clear separation of concerns: while the risk trajectory directly supports the active clinical workflow, the population analysis serves to support the broader scientific workflow.

\begin{custommdframed}
\textcolor{DeepPurple}{\textbf{\textit{RQ1's Key Findings and Implications:}}}
\vspace{0.2em}

\circlednum{1} \textbf{Clinicians prioritize AI's explanatory power over its predictive output.} The high utility ratings for interpretable features (risk trajectory, feature importance), especially among junior clinicians, underscore a demand for tools that support, rather than bypass, clinical reasoning.

\circlednum{2} \textbf{Usability is highly context-dependent and challenged by information density.} The system's moderate overall SUS score, suggests that a one-size-fits-all design is insufficient. A primary design challenge is managing information density to avoid cognitive friction.

\circlednum{3} \textbf{Generative recommendations require strict grounding.} The LLM module was valued for efficiency but criticized for lack of domain specificity and occasional hallucination. To be viable, generative features in high-stakes settings must move beyond general summarization to guideline-anchored and strictly verifiable output.
\end{custommdframed}

\subsection{Analysis to RQ2}

\paragraph{AICare did not statistically alter overall efficiency, but qualitative patterns suggest expertise-based usage differences.}

Our study reveals a nuanced impact of AICare on diagnostic efficiency. To capture precise interaction behaviors and timing, we relied on high-resolution screen recordings. Valid screen recordings were available for a subset of 12 participants (9 senior clinicians and 3 junior clinicians), determined by participant consent and data completeness. Consequently, the following analyses on efficiency and interaction metrics are restricted to this subgroup.

Overall, the average time taken to complete a case was comparable between the AI-assisted and unassisted conditions (Table~\ref{tab:task_completion_time}, $p=.774$). When stratifying participants by clinical experience, we observed a potential trend: junior clinicians ($\leq$ 5 years of experience) showed a decrease in average task time of approximately 32\%. Although this result did not reach statistical significance ($p=.130$) given the subgroup sample size ($N=3$), the substantial reduction suggests a distinct behavioral adaptation compared to senior clinicians ($>$ 5 years of experience), who took slightly longer when using the system.

\begin{table}[!ht]
    \centering
    \caption{Comparison of task completion time (in seconds) between the AI-assisted and unassisted conditions. Data analysis is based on $N=12$ participants (9 Seniors, 3 Juniors) for whom complete, valid screen recordings were available. Statistical significance was assessed using repeated-measures ANOVA (RM-ANOVA). Note that for the junior clinicians group, a large effect size (partial $\eta_p^2$ = 0.664) was observed, indicating a substantial difference in task duration despite the statistical constraints of the subgroup sample size ($N=3$).}
    \label{tab:task_completion_time}
\resizebox{\columnwidth}{!}{%
    \begin{tabular}{@{} l r c r c c c c @{}}
        \toprule
        & \multicolumn{2}{c}{\textbf{AI-assisted}} & \multicolumn{2}{c}{\textbf{Unassisted}} & \multicolumn{3}{c}{\textbf{Statistical Comparison}} \\
        \cmidrule(lr){2-3} \cmidrule(lr){4-5} \cmidrule(lr){6-8}
        \textbf{Group} & \textbf{N} & \textbf{Mean\stddev{SD}} & \textbf{N} & \textbf{Mean\stddev{SD}} & \textbf{F} & \textbf{p} & \textbf{$\eta_p^2$} \\
        \midrule
        \textbf{Overall} & 12 & 117.68\stddev{63.24} & 12 & 113.25\stddev{54.63} & 0.087 & .774 & 0.012 \\
        \midrule
        \textbf{By Department} & & & & & & & \\
        \quad XY-Nephrology & 8 & 134.57\stddev{67.71} & 8 & 124.85\stddev{57.80} & 0.601 & .463 & 0.041 \\
        \quad BC-Obstetrics & 3 & 95.87\stddev{40.83} & 3 & 96.07\stddev{53.74} & 0.250 & .667 & <.001 \\
        \midrule
        \textbf{By Seniority} & & & & & & & \\
        \quad Senior clinicians & 9 & 131.96\stddev{61.23} & 9 & 114.09\stddev{60.39} & 1.449 & .263 & 0.209 \\
        \quad Junior clinicians & 3 & 74.87\stddev{57.75} & 3 & 110.73\stddev{42.58} & 6.250 & .130 & 0.664 \\
        \bottomrule
    \end{tabular}%
}
\end{table}

To understand the behavioral mechanisms driving this temporal difference, we analyzed granular interaction logs (Table~\ref{tab:interaction_metrics}). The data uncovers a statistically significant disparity in verification strategies: senior clinicians engaged in substantially more rigorous data interrogation than juniors. Specifically, seniors performed significantly more curve hovering ($F=41.33, p<.001, \eta_p^2=0.805$) and cross-view comparison ($F=7.07, p=.024, \eta_p^2=0.414$). This indicates that experts did not passively accept the AI's risk prediction; rather, they actively manipulated the timeline to inspect specific data points and frequently moved their focus between historical trends and current evidence. In contrast, junior clinicians showed significantly lower interaction frequencies (e.g., mean list paging $0.93$ vs. $2.04$ for seniors). This behavioral divergence suggests distinct reliance strategies: while experts engaged in ``adversarial verification'' by physically interrogating the data to test the model's logic, junior clinicians appeared to utilize the system's pre-computed synthesis as a ``cognitive scaffold,'' relying on the provided visual structure to frame their analysis rather than performing granular, bottom-up validation.

\begin{table}[!ht]
\centering
\caption{Comparison of interaction behaviors between senior and junior clinicians. Data analysis is based on $N=12$ participants (9 Seniors, 3 Juniors) for whom complete, valid screen recordings were available. An asterisk (*) indicates a statistically significant result. ($p<.05$)}
\label{tab:interaction_metrics}
\resizebox{\columnwidth}{!}{%
\begin{tabular}{@{}l cc ccc@{}}
\toprule
& \multicolumn{1}{c}{\textbf{Senior}} & \multicolumn{1}{c}{\textbf{Junior}} & \multicolumn{3}{c}{\textbf{Statistical Comparison}} \\
& \multicolumn{1}{c}{$(N=9)$} & \multicolumn{1}{c}{$(N=3)$} & & & \\
\cmidrule(lr){2-2} \cmidrule(lr){3-3} \cmidrule(lr){4-6}
\textbf{Metric} & \textbf{Mean\stddev{(SD)}} & \textbf{Mean\stddev{(SD)}} & \textbf{\textit{F}} & \textbf{\textit{p}} & \textbf{$\eta_p^2$} \\
\midrule
List Paging & 2.04\stddev{(0.56)} & 0.93\stddev{(0.09)} & 9.615 & \textbf{.011$^{*}$} & 0.490 \\
Curve Hovering & 2.13\stddev{(0.28)} & 0.60\stddev{(0.43)} & 41.328 & \textbf{< .001$^{*}$} & 0.805 \\
Cross-View Comparison & 1.69\stddev{(0.69)} & 0.40\stddev{(0.57)} & 7.067 & \textbf{.024$^{*}$} & 0.414 \\
\bottomrule
\end{tabular}%
}
\end{table}

Regarding accuracy, we evaluated diagnostic performance on a subset of cases ($N=8$ participants from XY-Nephrology) where robust ground truth outcomes were available and the sample size allowed for valid comparison. As shown in Table~\ref{tab:diagnostic_accuracy}, while accuracy improved marginally in the AI-assisted condition (+10.0\%), the difference was not statistically significant ($p=.619$).

\begin{table}[!ht]
    \centering
    \caption{Comparison of diagnostic accuracy between AI-assisted and unassisted conditions across different clinician groups. Analysis is restricted to the XY-Nephrology cohort ($N=8$) where the sample size and ground truth distribution supported a valid comparative analysis. The AI model's benchmarked performance on the test set (10 cases from XY-Nephrology) was: Accuracy 80.0\%, Precision 66.7\%, Recall 66.7\%, Specificity 85.7\%, AUROC 0.643, and AUPRC 0.542. Repeated-measures ANOVA (RM-ANOVA) was used to compare accuracy between conditions. An improvement with a ``+'' indicates higher accuracy in the AI-assisted condition.}
    \label{tab:diagnostic_accuracy}
\resizebox{\columnwidth}{!}{%
    \begin{tabular}{@{} l c c c c c @{}}
        \toprule
        \textbf{Group} & \textbf{N} & \textbf{AI-assisted (\%)} & \textbf{Unassisted (\%)} & \textbf{$\uparrow$ (\%)} & \textbf{p} \\
        \midrule
        \textbf{Overall clinicians} & 8 & 72.5\stddev{21.2} & 62.5\stddev{29.2} & +10.0 & .619 \\
        \midrule
        \textbf{By Seniority} & & & & & \\
        \quad Senior clinicians & 5 & 80.0\stddev{20.0} & 56.0\stddev{32.9} & +24.0 & .253 \\
        \quad Junior clinicians & 3 & 60.0\stddev{20.0} & 73.3\stddev{23.1} & -13.3 & .580 \\
        \bottomrule
    \end{tabular}
}
\end{table}

Qualitative feedback corroborates that these interaction patterns stem from distinct cognitive strategies. Junior clinicians tended to use AICare as a cognitive scaffold that structured and accelerated their analytical process. For example, resident XY-07A noted, ``The overall risk curve is very important... I will first look at it to judge the patient's overall risk for the next year.'' Obstetricians BC-01A, when dealing with the specific, time-bound risk of preterm birth, reported that the tool was particularly efficient for ``screening scenarios in outpatient clinics'', where rapid risk stratification is paramount. For them, the system provided a clear starting point and a structured overview. Senior clinicians, conversely, engaged in a more deliberate dialogue with the AI, using the extra time for verification and criticism. For instance, BS-04B, a chief physician, spent considerable time analyzing the ``cluttered'' charts to scrutinize the system's logic against her own. This suggests that for experts, the AI introduces a ``second opinion'' verification step that, while valuable for safety, is inherently additive to the temporal workflow.

\paragraph{The system substantially reduced perceived cognitive workload across all clinician groups.}
A key benefit of AICare was its ability to alleviate the cognitive burden associated with analyzing complex patient cases. Participants reported a significant reduction in overall cognitive workload in the AI-assisted condition compared to the unassisted condition, as measured by the NASA-TLX scale (Table~\ref{tab:nasa_tlx_workload}, $p=.023$). This reduction was particularly pronounced among junior clinicians, who reported a statistically significant decrease in workload ($p=.028$).

\begin{table}[!ht]
    \centering
    \caption{Comparison of cognitive workload, measured by the NASA-TLX average score (scale 0-100), between the AI-assisted and unassisted conditions. Lower scores indicate a lower perceived workload. Statistical significance was assessed using repeated-measures ANOVA (RM-ANOVA). An asterisk (*) indicates a statistically significant result. ($p<.05$)}
    \label{tab:nasa_tlx_workload}
\resizebox{\columnwidth}{!}{%
    \begin{tabular}{@{} l r c r c c c c @{}}
        \toprule
        & \multicolumn{2}{c}{\textbf{AI-assisted}} & \multicolumn{2}{c}{\textbf{Unassisted}} & \multicolumn{3}{c}{\textbf{Statistical Comparison}} \\
        \cmidrule(lr){2-3} \cmidrule(lr){4-5} \cmidrule(lr){6-8}
        \textbf{Group} & \textbf{N} & \textbf{Mean\stddev{SD}} & \textbf{N} & \textbf{Mean\stddev{SD}} & \textbf{F} & \textbf{p} & \textbf{$\eta_p^2$} \\
        \midrule
        \textbf{Overall} & 16 & 41.55\stddev{15.54} & 16 & 47.49\stddev{12.51} & 6.385 & .023* & 0.182 \\
        \midrule
        \textbf{By Department} & & & & & & & \\
        \quad XY-Nephrology & 8 & 31.85\stddev{8.45} & 8 & 40.33\stddev{11.01} & 1.833 & .218 & 0.274 \\
        \quad BS-Nephrology & 5 & 52.13\stddev{16.92} & 5 & 55.70\stddev{11.43} & 4.500 & .101 & 0.220 \\
        \quad BC-Obstetrics & 3 & 49.78\stddev{15.21} & 3 & 52.89\stddev{8.40} & 0.333 & .622 & 0.041 \\
        \midrule
        \textbf{By Seniority} & & & & & & & \\
        \quad Senior clinicians & 9 & 38.35\stddev{14.12} & 9 & 44.39\stddev{12.73} & 1.195 & .306 & 0.134 \\
        \quad Junior clinicians & 7 & 45.67\stddev{17.40} & 7 & 51.48\stddev{11.92} & 8.292 & .028* & 0.370 \\
        \bottomrule
    \end{tabular}%
}
\end{table}

The system achieves this by offloading the demanding task of synthesizing and interpreting large volumes of longitudinal EHR data. The integrated visualizations, such as the dynamic risk trajectory (Figure~\ref{fig:aicare_platform}A) and the interactive feature importance list (Figure~\ref{fig:aicare_platform}B), provide a pre-digested analytical summary. This allows clinicians to focus their cognitive resources on higher-level clinical reasoning, interpretation, and verifying the AI's logic, rather than on the manual and time-consuming process of data aggregation. Participant BS-03A described the visualization of risk trends as ``very intuitive'', contrasting it with the high mental effort of reading tabular data. Even among senior clinicians who were critical of the interface's visual density, the reduction in mental demand was acknowledged.

\begin{custommdframed}
\textcolor{DeepPurple}{\textbf{\textit{RQ2's Key Findings and Implications:}}}
\vspace{0.2em}

\circlednum{1} \textbf{Efficiency impact is experience-dependent with maintained accuracy, driven by distinct interaction behaviors.} While overall time remained stable, interaction logs reveal that senior clinicians engage in significantly more active verification than juniors. This suggests AICare acts as a cognitive scaffold for novices but as a peer for experts.

\circlednum{2} \textbf{Cognitive load is significantly reduced.} By synthesizing complex longitudinal data into an intuitive dashboard, AICare offloads the cognitive burden of data aggregation. This frees up clinicians' mental resources to focus on higher-level patient care and complex decision-making.
\end{custommdframed}

\subsection{Analysis to RQ3}

\paragraph{Trust is actively constructed through verification, driving decision-making confidence.}

Overall, clinicians rated AICare with a moderately high degree of trust ($M=5.18/7$, $SD=0.78$; see Table~\ref{tab:trust_in_ai}). However, our qualitative findings indicate that clinician trust is not a static attribute passively granted to the AI, but an active process of verification. Quantitative results show that while diagnostic accuracy did not significantly improve, clinicians' decision-making confidence increased significantly ($p=.018$, Table~\ref{tab:diagnostic_confidence}).

\begin{table}[!ht]
    \centering
    \caption{Descriptive statistics for the trust in AI score, measured on a 7-point Likert scale, across different participant groups. An independent samples t-test between the ``AI First'' and ``Human First'' groups found no significant difference in trust scores. ($t(14) = 0.016, p = .987, d = 0.008$)}
    \label{tab:trust_in_ai}
    \begin{tabular}{@{} l l c c @{}}
        \toprule
        \textbf{Category} & \textbf{Group} & \textbf{N} & \textbf{Mean\stddev{SD}} \\
        \midrule
        \textbf{Overall} & All Participants & 16 & 5.18\stddev{0.78} \\
        \midrule
        \textbf{By Department} & & & \\
        & XY-Nephrology & 8 & 5.54\stddev{0.57} \\
        & BS-Nephrology & 5 & 5.07\stddev{0.77} \\
        & BC-Obstetrics & 3 & 4.42\stddev{0.88} \\
        \midrule
        \textbf{By Seniority} & & & \\
        & Senior clinicians & 9 & 5.37\stddev{0.95} \\
        & Junior clinicians & 7 & 4.94\stddev{0.42} \\
        \midrule
        \textbf{By AI Usage Order} & & & \\
        & AI First & 9 & 5.19\stddev{0.77} \\
        & Human First & 7 & 5.18\stddev{0.84} \\
        \bottomrule
    \end{tabular}
\end{table}

\begin{table}[!ht]
    \centering
    \caption{Comparison of clinician diagnostic confidence between the AI-assisted and unassisted conditions. Confidence was measured on a 5-point Likert scale (1 = Not confident at all, 5 = Very confident), where higher scores indicate greater confidence. Repeated-measures ANOVA (RM-ANOVA) was used to compare scores between the two conditions. An asterisk (*) indicates a statistically significant result. ($p < .05$)}
    \label{tab:diagnostic_confidence}
\resizebox{\columnwidth}{!}{%
    \begin{tabular}{@{} l r c r c c c c @{}}
        \toprule
        & \multicolumn{2}{c}{\textbf{AI-assisted}} & \multicolumn{2}{c}{\textbf{Unassisted}} & \multicolumn{3}{c}{\textbf{Statistical Comparison}} \\
        \cmidrule(lr){2-3} \cmidrule(lr){4-5} \cmidrule(lr){6-8}
        \textbf{Group} & \textbf{N} & \textbf{Mean\stddev{SD}} & \textbf{N} & \textbf{Mean\stddev{SD}} & \textbf{F} & \textbf{p} & \textbf{$\eta_p^2$} \\
        \midrule
        \textbf{Overall} & 16 & 3.71\stddev{0.77} & 16 & 3.29\stddev{0.87} & 7.018 & .018* & 0.253 \\
        \midrule
        \textbf{By Department} & & & & & & & \\
        \quad XY-Nephrology & 8 & 3.90\stddev{0.67} & 8 & 3.63\stddev{0.65} & 5.453 & .052 & 0.336 \\
        \quad BS-Nephrology & 5 & 3.72\stddev{0.80} & 5 & 3.00\stddev{1.17} & 2.087 & .222 & 0.284 \\
        \quad BC-Obstetrics & 3 & 3.20\stddev{1.06} & 3 & 2.87\stddev{0.76} & 4.000 & .184 & 0.641 \\
        \midrule
        \textbf{By Seniority} & & & & & & & \\
        \quad Senior clinicians & 9 & 3.73\stddev{0.89} & 9 & 3.51\stddev{0.81} & 6.400 & .035* & 0.347 \\
        \quad Junior clinicians & 7 & 3.69\stddev{0.65} & 7 & 3.00\stddev{0.92} & 4.703 & .073 & 0.324 \\
        \bottomrule
    \end{tabular}%
}   
\end{table}

Qualitative feedback reveals that this boost stems from the system's scrutability, which allows users to validate their own intuitions against the AI's evidence. Chief physician XY-08B described this dynamic as ``a process of verification and comparison'', where she used the tool not to generate an answer, but to confirm if her line of thought was consistent with the data. This verification was often facilitated by specific interactive mechanisms; as attending clinician XY-03A explained, ``When I click on a point in time, the system immediately shows all the key indicator values... This is very useful, it makes everything clear at a glance, and that enhances trust.'' This suggests that transparency was functionally defined by participants as verifiability. Trust could be reinforced when interactive features allowed clinicians to confirm that the AI's logic was grounded in clinically sound evidence, specifically the alignment of feature importance with established pathophysiological mechanisms. Similarly, resident BS-03A noted that she approached the system with a ``natural sense of trust'', but it was the visual confirmation of the risk trajectory that solidified her assurance. Consequently, even when the AI's prediction differed slightly from the clinician's initial view, this transparency reduced subjective uncertainty. As obstetrician BC-03A noted, the ``system's explanation function helps in understanding the difference'', enabling her to make a final decision with conviction. Thus, AICare helps transform the black box of prediction into a verifiable process, allowing clinicians to decouple their confidence from the model's raw accuracy.

\paragraph{Transparency is a double-edged sword: Plausibility determines trust.}
The interactive nature of AICare magnified the impact of the AI's reasoning quality. When the system's logic was transparent but divergent from clinical expectation, it often sparked productive negotiation. For example, one attending clinician (XY-03A) noted that while she typically focused on indicators like albumin and potassium, the AI's emphasis on sodium and PTH ``opened up my thinking.'' She explained that such a discrepancy would prompt her to investigate the literature, thereby using the AI as a tool to challenge her own ``fixed and limited'' diagnostic habits. However, this same transparency made trust highly fragile when the revealed logic was clinically implausible. Because clinicians could see the ``why'', errors were immediately glaring. Chief physician BS-04B expressed explicit distrust when the system prioritized Cystatin C in a context contradicting conventional knowledge. Associate chief nurse XY-06B also expressed skepticism regarding the model's feature weighting, explicitly stating she doubted whether ``the model might be overemphasizing'' the importance of uric acid across different patient contexts. More critically, XY-08B highlighted that logical contradictions, such as the LLM flagging ``digestive system disease'' when the feature value was zero, were fatal to trust: ``When the conclusion contradicts clinical common sense, the doctor cannot trust it.'' Furthermore, concerns about data integrity, such as BC-01A's skepticism regarding imputation methods for missing values, surfaced precisely because the system made these data dependencies visible. This demonstrates that while interpretability can bridge the gap between human and AI, it requires the AI's reasoning to be not just explainable, but clinically robust; otherwise, visibility accelerates the loss of trust. Furthermore, the boundaries of trust were reinforced by what resident XY-07A termed AI's ``observational blind spots'', where the system cannot capture non-quantifiable bedside data, such as subtle changes in mental state or edema, requiring the clinician to bridge the gap between algorithmic probability and patient reality.

\begin{custommdframed}
\textcolor{DeepPurple}{\textbf{\textit{RQ3's Key Findings and Implications:}}}
\vspace{0.2em}

\circlednum{1} \textbf{Verification drives confidence.} Clinicians do not blindly trust the AI; they test it. The significant rise in diagnostic confidence stems from the ability to interactively verify the AI's evidence (e.g., via drill-down interactions), which validates their own clinical intuition.

\circlednum{2} \textbf{Interpretability as a double-edged sword.} Transparency amplifies both trust and distrust. While it allows clinicians to catch errors (safety), it also means that ``hallucinations'' or illogical feature weights (e.g., zero-value risks) cause immediate, catastrophic drops in trust.

\circlednum{3} \textbf{Plausibility enables learning.} Disagreement between Human and AI fosters learning only when the AI's rationale is clinically plausible. Without grounding in standard medical logic or external evidence, novel insights are dismissed as algorithmic artifacts.

\circlednum{4} \textbf{Data integrity and observational blind spots delimit trust.} Transparency regarding data dependencies exposes vulnerabilities that can erode trust. Furthermore, clinicians explicitly withhold trust for non-quantifiable physical signs and subjective patient states that remain invisible to the model.
\end{custommdframed}

\section{Discussion}

\begin{figure*}[!ht]
\centering
\includegraphics[width=0.9\linewidth]{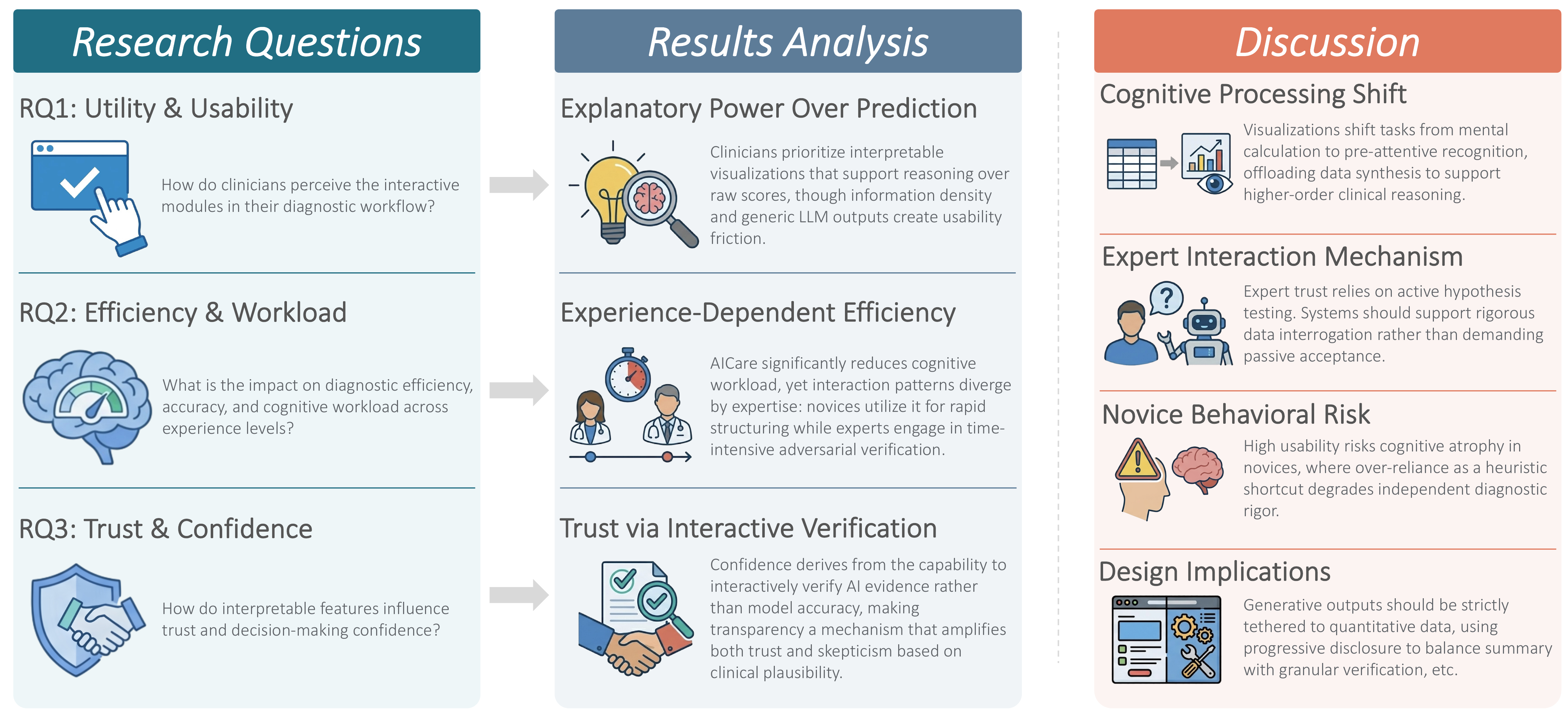}
\caption{Study framework mapping the progression from research questions to results and discussion. The diagram illustrates three core pathways: the shift from prediction to explanatory power (RQ1), the divergence of interaction strategies based on clinical expertise (RQ2), and the active construction of trust through verification (RQ3), culminating in implications for cognitive processing and system design.}
\label{fig:aicare_results}
\end{figure*}

We interpret our findings by synthesizing the quantitative results from the primary study with the narrative insights derived from the supplementary interview cohort ($N=4$, described in Table~\ref{tab:more_participants}). Specifically, we discuss the shift from tabular cognition to visual recognition, the divergence in verification behaviors between experts and novices, and the resulting implications for designing practical clinical AI. As synthesized in Figure~\ref{fig:aicare_results}, our investigation follows a progression from inquiry to implication.

\subsection{Does AICare Really Outperform Tabular Systems? The Role of Interactive Verification}

AICare outperforms the tabular status quo not merely by providing predictions, but by altering the cognitive architecture of the risk assessment task. Quantitatively, this is evidenced by a significant reduction in perceived cognitive workload compared to baseline tabular systems, with improvements ranging from 11.3\% to 36.1\% across participants (BC-A1: 32.3\%, BC-A2: 36.1\%, BC-A3: 11.3\%, BC-A4: 12.5\%). Theoretical frameworks in cognitive engineering and medical informatics suggest that standard hospital information systems force clinicians to perform high-effort ``mental simulation'' on matrices of static values to reconstruct patient history~\cite{manning2013visualizing, zhang2024rethinking}. By externalizing this computation into the dynamic risk trajectory, AICare replaces the memory-intensive task of integrating row-and-column data with a high-bandwidth visual channel. This aligns with findings by Zaj{\k{a}}c et al., who argue that clinically useful AI must move beyond accuracy to support the temporal and contextual realities of workflow integration~\cite{zajkac2025towards}. Consequently, clinicians can perceive critical trends (e.g., deterioration slopes) as pre-attentive attributes rather than calculated derivatives.

Qualitatively, the superiority of the AI copilot stems from its ability to mitigate the information overload inherent in high-dimensional tabular datasets~\cite{kushniruk2004cognitive}. The cognitive burden of raw matrices was explicitly highlighted, as standard systems make it ``difficult to process every detail'' (BC-A2). By directly rendering the ``trend of indicators'' (BC-A1), the system replaces memory-intensive search with pre-attentive recognition. This shift transforms verification from a disjointed task into an immediate investigation; whereas tabular interfaces force users to hunt for correlations, the copilot makes the specific ``contribution'' of each feature accessible on demand (BC-A3). Furthermore, to mitigate the risk of missing signals in dense data, the system enables users to ``verify AI warnings'' by quickly locating indicator changes in key gestational weeks, acting as a safety net for subtle patterns (BC-A4)~\cite{wang2021brilliant}.

\subsection{Why and When Clinicians Challenge AI Reasoning?}

Clinicians, particularly experts, challenge AI reasoning because trust is constructed through a process of adversarial verification rather than passive acceptance. Our interaction metrics revealed that senior clinicians engaged in high-frequency data interrogation (e.g., curve hovering) not due to inefficiency, but because they treat AI outputs as hypotheses requiring rigorous evidence. This behavior mirrors the ``negotiation'' phase described by Sivaraman et al., where clinicians actively probe AI outputs to establish reliability before acceptance~\cite{sivaraman2023ignore}. Qualitative feedback elucidates that this scrutiny arises from a need to map AI logic onto internal mental models; trust is effectively established only when the feature importance ``conforms to medical logic'' (BC-A3), a phenomenon consistent with the ``contestability'' requirements for trust in clinical CDSS~\cite{tun2025trust}. Furthermore, static risk probabilities fail to satisfy expert requirements for causal explanation. Clinicians cannot judge urgency from a score alone without visualizing the ``change and slope'' of indicators (BC-A2). Therefore, the challenge behavior is a mechanism to expose the underlying evidence, transforming the AI from a black-box oracle into a scrutable peer that must demonstrate its reasoning process to be trusted. This aligns with frameworks on meaningful human oversight, where interpretable features act as instruments for verification rather than blind reliance~\cite{mcgrath2025collaborative, goh2025physician}. By enabling active auditing, the system preserves professional accountability, ensuring the clinician remains the responsible agent.

Active interrogation of AI outputs is triggered specifically during complex decision-making contexts involving ``grey area situations'' or safety-critical redundancy checks. While junior clinicians utilize the system as a cognitive scaffold to combat the ``mental inertia'' of processing massive datasets (BC-A2), senior clinicians activate their critical verification primarily when facing ambiguity. In optional treatment scenarios where the path forward is unclear, the system serves as a crucial ``second medical opinion'' (BC-A4), a role that Lu et al. found significantly improves decision quality when the AI provides distinct, evidence-based reasoning~\cite{lu2024does}. Conversely, in routine checks, the motivation shifts from decision support to safety assurance; the core value becomes ``checking for omissions and filling gaps'' rather than replacing human judgment (BC-A1). Thus, clinicians challenge the AI most intensely not when the case is obvious, but when the decision boundaries are blurred or when the cost of a missed indicator is unacceptably high~\cite{wolf2025clinical}.

\subsection{Risk of Algorithmic Over-reliance and Passive Acceptance in Junior Clinicians}

Our observations point to a potential vulnerability in the deployment of clinical AI systems: the susceptibility of novice clinicians to automation bias. While the system successfully mitigates the ``mental inertia'' caused by massive data overload, this cognitive relief creates a dangerous pathway toward passivity. For junior clinicians, who are still developing their internal diagnostic models, the high usability of AICare may inadvertently encourage them to treat the system's output as a verified conclusion rather than a hypothesis to be tested~\cite{gaube2021ai}.

The danger is that the AI functions not as a pedagogical scaffold, but as a heuristic shortcut, leading to premature closure where the clinician bypasses the essential step of independent verification. This dynamic threatens to fundamentally erode the diagnostic rigor required for high-stakes medicine. By viewing the AI primarily as a ``magical tool'' for efficiency rather than a peer for consultation, junior clinicians risk a gradual de-skilling of their critical faculties. If the visualization of risk trajectories is accepted without scrutiny, the clinician is reduced from an active analyst to a passive supervisor, creating a scenario where they are liable for algorithmic hallucinations they are no longer equipped to catch~\cite{klingbeil2024trust}. Consequently, the very features designed to augment clinical capacity may paradoxically atrophy the critical thinking necessary for safe, independent practice unless deliberate cognitive forcing functions are introduced~\cite{buccinca2021trust}.

\subsection{Design Implications for Next Generation Clinical AI Copilots}

To bridge the gap between algorithmic output and clinical cognition, future systems could address distinct behavioral modes through the following strategies:

\begin{enumerate}[leftmargin=*, label=(\arabic*), topsep=0pt]
    \item \textbf{Strengthen the coupling between generative summaries and quantitative evidence.}
    To mitigate the risk of hallucinations and support the verification behaviors observed in our study, interfaces should tether LLM-generated narratives directly to their quantitative sources. For instance, future systems could make the provenance of each claim explicit by visually linking narrative points to the underlying data, instead of presenting the summary in isolation. This approach would enable clinicians to seamlessly validate the AI's qualitative interpretation against the ground-truth records.

    \item \textbf{Manage cognitive load through context-aware progressive disclosure.}
    To combat information overload without hiding critical data, systems should adopt a layered information architecture. Instead of presenting all feature contributions simultaneously, the interface should initially present only the ``synthesis layer'' (overall risk trend and top 3 drivers). Detailed feature lists and population analytics should be revealed only on demand or triggered by specific anomaly detections. For junior clinicians, this acts as cognitive scaffolding; for experts, it reduces visual noise, allowing them to drill down only when the synthesis contradicts their intuition.

    \item \textbf{Prioritize clinical sense-making to drive everyday acceptance.}
    While interactive verification builds initial trust, sustained acceptance in daily workflows relies on shifting the focus from reasoning about the AI to reasoning about the patient. Our findings suggest that if an AI system demands constant, high-friction auditing, it risks becoming a burden rather than a support. To ensure long-term adoption, design strategies must go beyond merely exposing feature weights. Instead, they must align with the clinician's natural sense-making process, which involves synthesizing fragmented data into a coherent patient narrative. By positioning the AI as a partner that actively supports the construction of the clinical picture rather than a logic puzzle to be solved, the system transitions from a liable tool to an accepted professional asset.

    \item \textbf{Treat algorithmic competence as a design prerequisite.}
    Interactive transparency acts as a magnifying glass: it builds trust when the model is robust but accelerates rejection when logic is flawed. Although techniques like retrieval-augmented generation (RAG) can mitigate hallucinations by grounding generative outputs in established clinical guidelines, they cannot compensate for fundamental algorithmic deficiencies. If the underlying retrieval algorithm fails to surface specific, high-confidence evidence, the design must suppress generative output rather than risking plausible but false fabrications. Ultimately, high-fidelity predictive performance remains the non-negotiable foundation upon which safe interaction is built.

\end{enumerate}

\subsection{Limitations, Future Work and Ethical Implications}

Our study has limitations characteristic of early-stage sociotechnical evaluations, specifically a modest sample size and reliance on retrospective data which constrain statistical generalizability. Technologically, the simultaneous visualization of high-dimensional longitudinal data introduced visual complexity, occasionally increasing cognitive load, while the probabilistic nature of LLMs poses safety risks regarding plausible yet unverified hallucinations. These findings highlight the need for optimized visual hierarchies and rigorous grounding mechanisms to balance information density with clinical safety.

Future work will leverage the ongoing deployment to conduct longitudinal analysis of usage patterns over extended periods. We aim to implement interactive counterfactual analysis where clinicians can manually modulate input variables, such as lab values, to immediately observe shifts in the risk trajectory, allowing doctors to stress-test clinical hypotheses directly. Additionally, we will develop adaptive interfaces that tailor guidance to user expertise, providing pedagogical scaffolding for novices and rapid verification for experts.

Broadly, AICare advances a paradigm shift from automation to cognitive augmentation. However, deploying such systems necessitates strict safeguards against automation bias and liability. Passive acceptance by novices risks atrophying independent diagnostic skills, while generative hallucinations threaten to burden clinicians with liability for unforeseen errors. Ethical design must therefore enforce cognitive friction to compel active verification, ensuring the system functions as a tool for scrutiny rather than a shortcut to premature closure. Ultimately, this work provides a blueprint for responsible AI integration where algorithmic intelligence empowers rather than displaces human judgment in high-stakes environments.

\section{Conclusion}

Our work reframes the challenge of medical AI adoption from one of algorithmic accuracy to sociotechnical collaboration. By evaluating AICare across distinct clinical specialties, we demonstrate that interactive interpretability empowers clinicians to actively verify rather than passively accept algorithmic risk assessments. The significant reduction in cognitive workload and the parallel increase in diagnostic confidence confirm that well-designed AI copilots can manage the information overload of longitudinal care without compromising human agency. Crucially, the distinct interaction patterns observed between experts and novices reveal that transparent systems could serve dual roles: functioning as platforms for adversarial verification by senior clinicians to validate model behavior, and as pedagogical scaffolds for junior clinicians to structure their reasoning. We conclude that the future of responsible healthcare lies in integrated systems that prioritize scrutability over simple automation, fostering a symbiotic partnership where computational power amplifies rather than replaces the irreplaceable nuance of clinical expertise.

\begin{acks}
This work was supported by National Natural Science Foundation of China (62402017, 82470774, 82561128248, 82504426), Research Grants Council of Hong Kong (27206123, 17200125, C5055-24G, T45-401/22-N), Hong Kong Innovation and Technology Fund (GHP/318/22GD), Beijing Natural Science Foundation (QY25224, L244063, L244025), Peking University (Clinical Medicine Plus X Pilot Project 2024YXXLHGG007; ``TengYun'' Clinical Research Program TY2025015). Liantao Ma was supported by Beijing Traditional Chinese Medicine Science and Technology Development Fund (BJZYZD-2025-13), Beijing Municipal Health Commission Research Ward Excellence Clinical Research Program (BRWEP2024W032150205), and Xuzhou Scientific Technological Projects (KC23143). Junyi Gao acknowledges the receipt of studentship awards from the Health Data Research UK-The Alan Turing Institute Wellcome PhD Programme in Health Data Science (grant 218529/Z/19/Z) and Baidu Scholarship. We also thank the Apple Education Team for their support in hardware environment deployment.
\end{acks}

\bibliographystyle{ACM-Reference-Format}
\bibliography{ref}

\appendix

\section{Data Inclusion and Cohort Selection}
\label{app:cohort_selection}

This study utilizes retrospective datasets derived from the EHR data of three departments. To ensure data privacy and integrity, we de-identify all personally identifiable information (PII) prior to analysis.

\subsection{Nephrology Cohorts Selection}

\begin{figure}[!ht]
  \centering
  \includegraphics[width=1.0\linewidth]{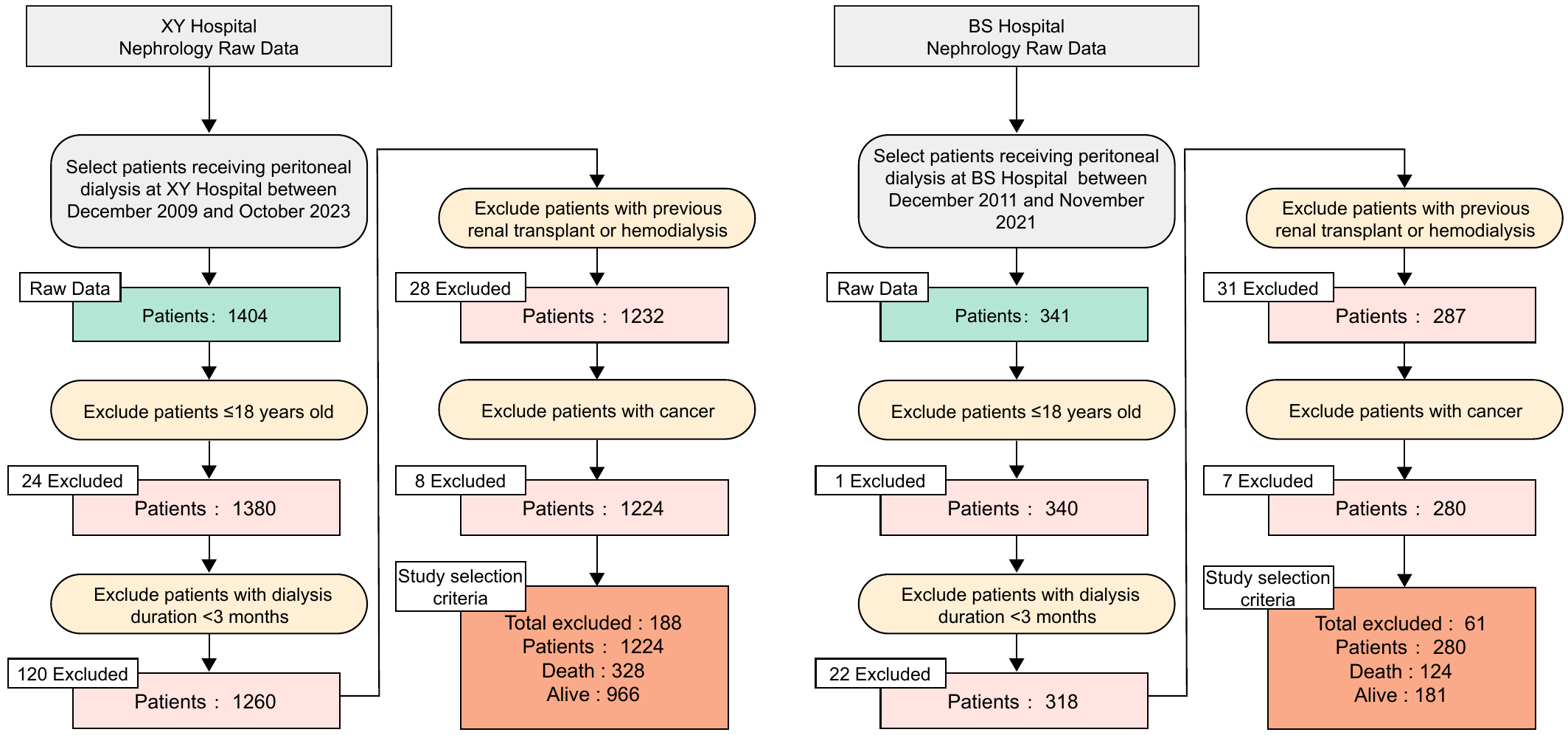} 
  \caption{The flowchart of participant selection for nephrology cohorts. The left panel displays the screening process for XY Hospital, while the right panel depicts the process for BS Hospital. The procedure sequentially filters patients based on age, dialysis vintage, history of other renal therapies, and presence of malignant tumors.}
  \label{fig:nephrology_inclusion_exclusion_flowchart}
\end{figure}

We establish retrospective cohorts focusing on patients undergoing Peritoneal Dialysis (PD) from two medical centers: XY Hospital (covering December 2009 to October 2023) and BS Hospital (covering December 2011 to November 2021). To ensure the study population represents a stable adult PD cohort and to minimize confounding factors from other treatments or severe comorbidities, we apply identical inclusion and exclusion criteria across both institutions. Figure~\ref{fig:nephrology_inclusion_exclusion_flowchart} illustrates the detailed participant selection process for both datasets.

Patients are included in the final cohorts if they meet the following criteria: (1) Adult status: The patient is older than 18 years at the time of data entry. This criterion excludes pediatric cases, which often require distinct clinical management strategies. (2) Stable dialysis vintage: The patient has maintained regular peritoneal dialysis for a duration of at least three months. This period allows for the stabilization of physiological parameters following the initial catheterization and dialysis induction.

To ensure the homogeneity of the dataset regarding the specific effects of peritoneal dialysis, we exclude patients based on the following conditions: (1) History of alternative renal replacement therapies: We exclude patients with a documented history of kidney transplantation or those who underwent emergency hemodialysis prior to PD. These treatments introduce physiological variations that differ significantly from those solely on peritoneal dialysis. (2) Malignant comorbidities: We exclude patients diagnosed with malignant tumors or cancer, as the systemic impact of malignancy and associated treatments (e.g., chemotherapy) significantly alters metabolic profiles and survival outcomes, potentially confounding the analysis of dialysis-related factors.

\subsection{Obstetrics Cohorts Selection}

\begin{figure}[!ht]
  \centering
  \includegraphics[width=1.0\linewidth]{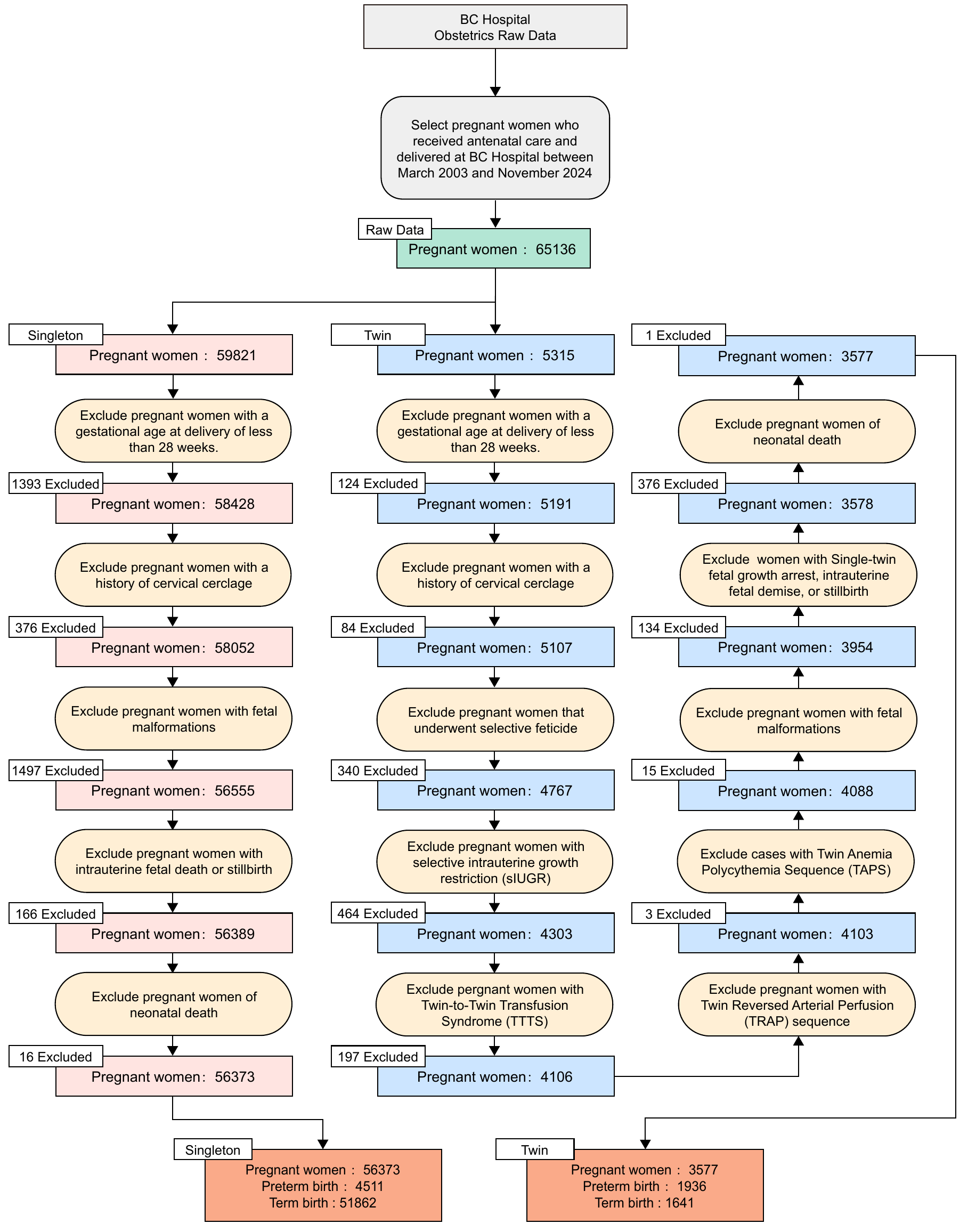} 
  \caption{Participant selection flowchart for the obstetrics dataset. The workflow distinguishes between singleton and twin pregnancies, sequentially applying exclusion criteria regarding gestational age, medical history, and fetal complications.}
  \label{fig:obstetrics_inclusion_exclusion_flowchart}
\end{figure}

We obtained obstetrics data from the EHR system of BC Hospital, covering the period from March 2003 to November 2024. Figure~\ref{fig:obstetrics_inclusion_exclusion_flowchart} illustrates the detailed participant selection process. In this study, we focused on the twin cohort.

We applied specific exclusion criteria to strictly define the study population and minimize confounding variables. For the singleton cohort, we excluded women based on the following conditions: (1) gestational age at delivery of less than 28 weeks, (2) a history of cervical cerclage, (3) fetal malformations, (4) intrauterine fetal death or stillbirth, and (5) neonatal death. For the twin cohort, the exclusion criteria further addressed complications specific to multiple gestations. In addition to gestational age less than 28 weeks and history of cervical cerclage, we excluded cases involving: (1) selective feticide, (2) selective intrauterine growth restriction (sIUGR), (3) twin-to-twin transfusion syndrome (TTTS), (4) twin anemia-polycythemia sequence (TAPS), (5) twin reversed arterial perfusion (TRAP) sequence, (6) fetal malformations, (7) single-twin fetal growth arrest, intrauterine fetal demise, or stillbirth, and (8) neonatal death. Following this selection process, the final dataset comprised 56,373 singleton pregnancies and 3,577 twin pregnancies.

\section{Dataset Features Statistics}
\label{app:dataset_features_stats}

This appendix presents the comprehensive statistics for the clinical features included in the three datasets used in this study, as detailed in Table~\ref{tab:xy_features} (XY Hospital), Table~\ref{tab:bs_features} (BS Hospital), and Table~\ref{tab:bc_features} (BC Hospital).

\begin{table}[!ht]
\centering
\footnotesize
\caption{Static and dynamic features statistics for the XY Hospital dataset ($N=1,224$). Complete list of static ($N=7$) and dynamic ($N=33$) features for the XY Hospital dataset (Nephrology). Units for binary or categorical variables are denoted as `-'.}
\label{tab:xy_features}
\resizebox{\columnwidth}{!}{%
\begin{tabular}{@{}lclclc@{}}
\toprule
\textbf{Feature} & \textbf{Unit} & \textbf{Feature} & \textbf{Unit} & \textbf{Feature} & \textbf{Unit} \\
\midrule
\multicolumn{6}{c}{\textbf{Static Features}} \\
\midrule
Height & cm & Chronic Nephritis/IgA & - & Polycystic Kidney & - \\
Weight & kg & Hypertensive Nephrop. & - & IgA Nephropathy & - \\
BMI & kg/m$^2$ & & & & \\
\midrule
\multicolumn{6}{c}{\textbf{Dynamic Features}} \\
\midrule
Real-time Age & years & Sodium & mmol/L & Total Protein & g/L \\
Dialysis Vintage & years & Uric Acid & $\mu$mol/L & Total Cholesterol & mmol/L \\
Albumin & g/L & Glucose & mmol/L & TIBC & $\mu$mol/L \\
WBC & $10^9$/L & Prealbumin & mg/L & Resp. Sys. Diseases & count \\
ALT & U/L & AST & U/L & PD-related Comp. & count \\
LDL-Cholesterol & mmol/L & Ferritin & ng/mL & Cardio-cereb. Comp. & count \\
Calcium & mmol/L & Hemoglobin & g/L & Digestive Sys. Dis. & count \\
Triglycerides & mmol/L & iPTH & pg/mL & Acute Upper Resp. & count \\
HDL-Cholesterol & mmol/L & Platelets & $10^9$/L & PD-related Perit. & count \\
MCV & fL & ALP & U/L & Creatinine & $\mu$mol/L \\
Potassium & mmol/L & Phosphorus & mmol/L & Chloride & mmol/L \\
\bottomrule
\end{tabular}
}
\end{table}

\begin{table}[!ht]
\centering
\scriptsize
\caption{Static and dynamic features statistics for the BS Hospital dataset ($N=280$). Complete list of static ($N=23$) and dynamic ($N=66$) features for the BS Hospital dataset (Nephrology). Units for binary or categorical variables are denoted as `-'.}
\label{tab:bs_features}
\resizebox{\columnwidth}{!}{%
\begin{tabular}{@{}lclclc@{}}
\toprule
\textbf{Feature} & \textbf{Unit} & \textbf{Feature} & \textbf{Unit} & \textbf{Feature} & \textbf{Unit} \\
\midrule
\multicolumn{6}{c}{\textbf{Static Features}} \\
\midrule
Gender & - & Pre-PD Albumin & g/L & Hypertensive Renal & - \\
Diabetes & - & Pre-PD Hemoglobin & g/L & Polycystic Kidney & - \\
BMI & kg/m$^2$ & Pre-PD Creatinine & $\mu$mol/L & Obstructive Nephrop. & - \\
Hist. CHD & - & Diabetic Nephrop. & - & HSP Nephritis & - \\
Height (at Dialysis) & cm & Chronic Interstitial & - & IgA Nephropathy & - \\
Weight (at Dialysis) & kg & Chronic Glomerulo. & - & Ischemic Nephrop. & - \\
Hist. Old MI & - & Hist. Cerebral Infarc. & - & Hist. Cerebral Bleed & - \\
Follow-up Peritonitis & - & Pre-PD Emergency HD & - & & \\
\midrule
\multicolumn{6}{c}{\textbf{Dynamic Features}} \\
\midrule
Real-time Age & years & HCT & \% & Neutrophil \% & \% \\
Dialysis Vintage & years & HDL-C & mmol/L & Phosphorus (P) & mmol/L \\
Albumin (ALB) & g/L & Hemoglobin (HGB) & g/L & P-LCR & \% \\
ALP & U/L & Potassium (K) & mmol/L & Prealbumin (PAB) & mg/L \\
ALT & U/L & LDH & U/L & PCT & \% \\
AST & U/L & LDL-C & mmol/L & PDW & fL \\
Basophil \# & $10^9$/L & Lymphocyte \# & $10^9$/L & Platelets (PLT) & $10^9$/L \\
Basophil \% & \% & Lymphocyte \% & \% & RBC & $10^{12}$/L \\
C1q & mg/L & MCH & pg & RDW-CV & \% \\
CK & U/L & MCHC & g/L & RDW-SD & fL \\
CK-MB & U/L & MCV & fL & Total Bilirubin & $\mu$mol/L \\
Chloride (Cl) & mmol/L & Monocyte \# & $10^9$/L & Total Cholesterol & mmol/L \\
Calcium (Ca) & mmol/L & Monocyte \% & \% & TCO2 & mmol/L \\
Creatinine (Cr) & $\mu$mol/L & MPV & fL & Triglycerides (TG) & mmol/L \\
Direct Bilirubin & $\mu$mol/L & Sodium (Na) & mmol/L & Total Protein (TP) & g/L \\
Eosinophil \# & $10^9$/L & Neutrophil \# & $10^9$/L & Uric Acid (UA) & $\mu$mol/L \\
Eosinophil \% & \% & Urea & mmol/L & US-CRP & mg/L \\
Globulin (GLB) & g/L & WBC & $10^9$/L & Cystatin C & mg/L \\
HBDH & U/L & $\beta$2-MG & mg/L & $\gamma$-GT & U/L \\
MLR & ratio & PLR & ratio & MPV/PC Ratio & ratio \\
Resp. Sys. Diseases & count & Cardio-cereb. Dis. & count & Acute Upper Resp. & count \\
PD-related Comp. & count & Digestive Sys. Dis. & count & PD-related Perit. & count \\
\bottomrule
\end{tabular}
}
\end{table}

\begin{table}[!ht]
\centering
\scriptsize
\caption{Static and dynamic features statistics for the BC Hospital dataset ($N=3,577$, Twin Cohort). Complete list of static ($N=29$) and dynamic ($N=50$) features for the BC Hospital dataset (Obstetrics). Units for binary or categorical variables are denoted as `-'.}
\label{tab:bc_features}
\resizebox{\columnwidth}{!}{%
\begin{tabular}{@{}lclclc@{}}
\toprule
\textbf{Feature} & \textbf{Unit} & \textbf{Feature} & \textbf{Unit} & \textbf{Feature} & \textbf{Unit} \\
\midrule
\multicolumn{6}{c}{\textbf{Static Features}} \\
\midrule
Maternal Age & years & Hist. C-Section & - & Preeclampsia & - \\
Height & cm & Family Hist. HTN & - & Preg. w/ Fibroids & - \\
Pre-pregnancy Weight & kg & Family Hist. Diabetes & - & Preg. w/ Adenomyosis & - \\
Hist. HTN & - & sIUGR & - & Chorionicity & - \\
Hist. Diabetes & - & GDM (Current) & - & PCOS & - \\
Hist. GDM & - & Pre-gestational Diab. & - & Nephrotic Syndrome & - \\
Hist. Preterm Birth & - & Gestational HTN & - & AKI & - \\
Hist. PROM & - & Pyelonephritis & - & SLE & - \\
Hist. Fibroids & - & OSA & - & APS & - \\
Recurrent Miscarriage & - & Sleep Breathing Dis. & - &  &  \\
\midrule
\multicolumn{6}{c}{\textbf{Dynamic Features}} \\
\midrule
Maternal Real-time Age & years & Total Protein (TP) & g/L & Creatinine (Cr) & $\mu$mol/L \\
Cervical Length & cm & Albumin (ALB) & g/L & Uric Acid (UA) & $\mu$mol/L \\
PT & s & Globulin (GLB) & g/L & CO2 & mmol/L \\
APTT & s & Total Bilirubin (TB) & $\mu$mol/L & Sodium (Na) & mmol/L \\
Thrombin Time (TT) & s & Direct Bilirubin (DB) & $\mu$mol/L & Potassium (K) & mmol/L \\
Fibrinogen (Fib) & g/L & Total Bile Acid & $\mu$mol/L & Chloride (Cl) & mmol/L \\
D-Dimer & mg/L & Urea & mmol/L & Calcium (Ca) & mmol/L \\
APTT Ratio (PAPTT) & ratio & Phosphorus (P) & mmol/L & TSH & mIU/L \\
PT Ratio (PTT) & ratio & Fasting Plasma Glu & mmol/L & Thyroxine (T4) & $\mu$g/dL \\
ALT & U/L & Free Thyroxine (FT4) & ng/dL & Free T3 (FT3) & pmol/L \\
AST & U/L & WBC & $10^9$/L & Neutrophils (NE) & $10^9$/L \\
ALP & U/L & Monocytes (MO) & $10^9$/L & Basophils (BAS) & $10^9$/L \\
Eosinophils (EOS) & $10^9$/L & Neutrophil \% & \% & Monocyte \% & \% \\
Basophil \% & \% & Eosinophil \% & \% & RBC & $10^{12}$/L \\
Hemoglobin (Hb) & g/L & MCV & fL & RDW-CV & \% \\
RDW-SD & fL & Platelets (PLT) & $10^9$/L & Mean Platelet Vol & fL \\
Plateletcrit (PCT) & \% & PDW & fL &  &  \\
\bottomrule
\end{tabular}
}
\end{table}

\section{Detailed Data Preprocessing and EHR Model Training}
\label{app:ehr_preprocessing_training}

This section provides a comprehensive description of the data processing, model architecture, training, and calibration procedures used to develop the predictive models for AICare. All procedures were implemented using our PyEHR toolkit~\cite{zhu2024pyehr}.

\subsection{Data preprocessing for nephrology.}
The pipeline for the end-stage renal disease (ESRD) datasets (XY and BS Hospitals) transforms raw patient records into a structured format suitable for time-series modeling. The key steps include:
\begin{itemize}[leftmargin=*, topsep=0pt]
    \item \textbf{Data integration and cleaning.} Records from multiple sources were merged and patient identifiers were standardized. Temporal data, such as visit and birth dates, were validated and harmonized.
    \item \textbf{Feature engineering.} Static features (e.g., primary renal disease, height) and dynamic features (e.g., laboratory results, comorbidities) were separated. We computed Body Mass Index (BMI), patient age, and dialysis vintage (duration of treatment) at each clinical visit. Categorical features were one-hot encoded based on clinical keywords.
    \item \textbf{Cohort definition and labeling.} The predictive task was to forecast mortality within a 365-day window. To prevent immortal time bias, we excluded all records for surviving patients that occurred within 365 days of their final follow-up~\cite{weiss2014incorporating}. A binary label was then assigned to each visit: ``1'' for mortality within 365 days and ``0'' otherwise.
    \item \textbf{Imputation and normalization.} We used a 10-fold stratified cross-validation setup. To prevent data leakage, imputation and normalization statistics were derived solely from the training set of each fold. Missing values were addressed first with a forward-fill imputation for each patient's timeline, followed by imputation of any remaining initial missing values using the training set median. All features were then Z-score normalized.
    \item \textbf{Class imbalance.} To mitigate the effects of an imbalanced class distribution, we applied random oversampling to the minority class within the training set of each fold.
\end{itemize}

\subsection{Data preprocessing for obstetrics.}
The pipeline for the obstetrics dataset (BC Hospital) predicting spontaneous preterm birth was largely consistent with the nephrology pipeline, with adaptations for the specific clinical context:
\begin{itemize}[leftmargin=*, topsep=0pt]
    \item \textbf{Data integration and identifier creation.} Data from outpatient, inpatient, laboratory, and demographic sources were integrated. A unique patient-pregnancy identifier was created to distinguish between multiple pregnancies for the same individual. Only records within 300 days prior to delivery were retained.
    \item \textbf{Cohort refinement.} A rigorous exclusion process was applied to create a homogeneous cohort. We removed cases with missing critical outcomes (e.g., gestational age at delivery) and those with clinical confounders such as cervical cerclage, selective fetal reduction, or fetal growth restriction.
    \item \textbf{Feature engineering and aggregation.} Static features (e.g., maternal age at delivery) and dynamic, time-varying features (e.g., gestational age at each visit) were created. To ensure a consistent time-series, multiple measurements within a single day were aggregated by their mean value. Features with a missingness rate above 90\% were pruned.
    \item \textbf{Fold-specific preprocessing.} A 10-fold stratified cross-validation was performed at the patient level to ensure all records from a patient-pregnancy dyad remained in the same data split. Imputation and normalization followed the same leakage-prevention protocol as in the nephrology pipeline, using a forward-fill strategy followed by median imputation for dynamic features and Z-score normalization based on training set statistics.
\end{itemize}

\subsection{Model architecture and training.}
Our AICare model (adapted from ConCare), designed to model complex interactions in longitudinal EHR data~\cite{ma2020concare}. The model processes patient trajectories through three main stages. First, a series of parallel Gated Recurrent Units (GRUs), one for each feature channel, captures the unique temporal evolution of each clinical variable. Second, these feature-specific representations are fed into a Transformer encoder block, where a multi-head self-attention mechanism models time-aware, cross-feature interactions. To encourage diverse representations, a decorrelation loss is applied to the attention head outputs. Finally, a terminal attention mechanism aggregates the contextualized feature vectors at each timestep to generate a dynamic patient representation and a corresponding risk prediction for each visit.

The model was implemented in PyTorch and trained using the PyTorch Lightning framework. We used the Adam optimizer with gradient clipping (max norm of 1.0) for numerical stability. The primary training objective used a Binary Cross-Entropy loss. An early stopping strategy was employed, monitoring the Area Under the Precision-Recall Curve (AUPRC) on the validation set. Training was halted if the validation AUPRC failed to improve for a specified number of epochs (patience), and the model checkpoint with the highest AUPRC was saved for evaluation.

\subsection{Post-hoc calibration and thresholding.}
To enhance the clinical utility of the model's predictions, we implemented a two-stage post-hoc optimization process on the validation set after training was complete~\cite{walsh2017beyond}.
\begin{itemize}[leftmargin=*, topsep=0pt]
    \item \textbf{Probability calibration.} We used Temperature Scaling to calibrate the model's outputs. This method learns a single scalar parameter (temperature, T) to rescale the pre-sigmoid logits, ensuring that the resulting probabilities more accurately reflect the true likelihood of the outcome. The optimal temperature is found by minimizing the Binary Cross-Entropy loss on the validation dataset~\cite{huang2020tutorial}.
    \item \textbf{Optimal decision threshold selection.} Following calibration, we determined the optimal decision threshold for converting probabilities into discrete class labels. We systematically searched through 200 candidate thresholds (from 0.01 to 0.99) to identify the value that maximized the F-beta score on the validation set. This allows the model's classification boundary to be tuned to the specific clinical costs of false positives versus false negatives.
\end{itemize}

\subsection{Hyperparameter configuration.}
The key hyperparameters used for training the AICare models on the three datasets are detailed in Table~\ref{tab:appendix_hyperparameters}. These parameters were determined through a combination of standard practices and experimentation on the validation sets~\cite{lluis2021distributing}.

\begin{table}[!ht]
  \caption{Hyperparameters for the AICare models on the three study datasets. Final values were selected based on performance on the validation set. ``Cls.'' means classification.}
  \label{tab:appendix_hyperparameters}
  \resizebox{\columnwidth}{!}{%
  \begin{tabular}{@{}lccc@{}}
    \toprule
    \textbf{Hyperparameter} & \textbf{XY-Nephrology} & \textbf{BS-Nephrology} & \textbf{BC-Obstetrics} \\
    \midrule
    Task              & Cls.   & Cls.   & Cls.     \\
    Seed              & 42          & 42          & 42            \\
    Epochs            & 30          & 30          & 30            \\
    Patience (early stopping)          & 10          & 10          & 5             \\
    Batch size       & 32          & 16          & 128           \\
    Learning rate    & 0.001       & 0.001       & 0.01          \\
    Primary metric      & AUPRC       & AUPRC       & AUPRC         \\
    Hidden dimension        & 128         & 128         & 32            \\
    Output dimension       & 1           & 1           & 1             \\
    Dynamic feature dimension          & 33          & 66          & 50            \\
    Static feature dimension          & 7           & 23           & 29            \\
    \bottomrule
  \end{tabular}
  }
\end{table}

\section{Prompt for Generating Diagnostic Recommendations}
\label{app:llm_generate_advice_prompt}

\begin{PromptBox}
\textbf{System prompt:}

You are an experienced clinician with extensive medical knowledge and clinical diagnostic experience.
\\

You will receive a patient's electronic health record (EHR) data, an AI model's risk prediction result, and feature importance weights. Based on this information, please conduct a clinical analysis and provide diagnostic and decision-making recommendations.
\\

Analysis Requirements:

1. Focus on the examination values from the most recent visit and the features with high importance weights.

2. Use analytical reasoning to deduce the patient's physiological or biochemical pathophysiological state.

3. Systematically identify the appropriate clinical response.

4. Provide specific clinical advice without listing the patient's specific data (do not use concrete numerical values).

5. Ensure the response is detailed and substantial.

\vspace{2em}

\textbf{User prompt template:}

**Clinical Prediction Task**

\{task\_definition\}
\\

**Patient's Electronic Health Record Analysis**

AI Model Risk Prediction Result: \{risk\_value\}\%
\\

Feature Importance Weights (Key Factors Influencing Prediction):

\{key\_features\_list\}
\\

Patient's Complete Examination Values from the Last Visit:

\{all\_feature\_values\}
\\

**Clinical Analysis Request**

Based on the AI model's analysis results and the patient's EHR data above, please use clinical reasoning to analyze the patient's pathophysiological state and provide specific diagnostic and decision-making recommendations.
\end{PromptBox}

To generate high-quality, personalized clinical advice, we have carefully engineered the prompt for interacting with a large language model (LLM). The prompt is constructed from two parts: a static system prompt that defines the model's role, core objectives, and output constraints, and a dynamic user prompt template that structures the specific patient data and AI analysis results for the task at hand. This separation ensures consistent model behavior while allowing for flexible adaptation to different clinical tasks.

\section{Guidance Script for Participants}
\label{app:guidance_script}
\textit{This script provides a consistent introduction for researchers to guide clinician participants through the study's questionnaires and interviews.}

\textbf{Introduction.}
``Thank you once again for your invaluable contribution to our research on the AICare system. Your professional insights are essential for us to understand the practical implications and potential of this technology. Next, we will conduct experiments, including a few standardized questionnaires and a brief concluding interview. Your honest and detailed feedback is deeply appreciated.''

\textbf{Part 1: Participant background questionnaire.}
``First, we have a brief background questionnaire to understand your professional profile, including your specialty, clinical experience, and general familiarity with AI technology. This information is purely for data analysis and helps us interpret the results across different clinician groups. There are no right or wrong answers.''

\textbf{Part 2: Standardized assessment scales.}
``In this section, we will use three internationally recognized assessment scales. These instruments are standard in human-computer interaction research and allow us to quantitatively measure key aspects of your experience, such as perceived workload, system usability, and trust. Using these validated scales ensures our findings are objective and comparable to established benchmarks in the field.''

``I will briefly explain each scale:''
\begin{enumerate}[leftmargin=*, label=(\arabic*), topsep=0pt]
    \item ``\textbf{NASA Task Load Index (NASA-TLX): Measuring your cognitive workload.} This scale assesses the mental and physical effort required to complete the diagnostic tasks. It is a two-step process. First, for six dimensions like `Mental Demand' or `Temporal Demand,' you will mark a line to indicate your feeling, from low to high. Please note that you will complete this questionnaire twice: once after performing the diagnostic tasks without assistance, and again after using the AICare system.''
    \item ``\textbf{System Usability Scale (SUS): Measuring how easy the system is to use.} This scale directly measures whether you found the AICare system to be straightforward, convenient, and well-designed. There are 10 statements; please indicate your level of agreement with each. Note that some statements are phrased positively (e.g., `I thought the system was easy to use') and others are phrased negatively (e.g., `I found the system unnecessarily complex'). Please read each statement carefully.''
    \item ``\textbf{Trust in Automation Scale: Measuring your trust in the AI.} This scale assesses your confidence in the analysis and recommendations provided by AICare. Given that trust is the foundation of human-AI collaboration in medicine, your perspective here is vital. Similar to the SUS, please read each statement and choose the option on the 7-point scale that best represents your opinion.''
\end{enumerate}

``Finally, before we move to the interview, we want to distinguish between the system as a whole and its specific parts. Listed here are the four main functional modules of AICare (the risk trajectory, the factor list, the LLM recommendation, and the population-level analysis). Please rate how useful you found each specific feature for your diagnostic process on a scale of 1 to 5.''

\textbf{Part 3: Semi-structured interview.}
``To conclude our session, we will have a semi-structured interview lasting approximately 15 minutes. While the questionnaires provide valuable quantitative data, this conversation is our opportunity to understand the context and the `why' behind your ratings. We are especially interested in your thought processes, specific experiences with the system, and your valuable clinical perspective.''

\section{Study Instruments}
This appendix contains all questionnaires and the interview guide used in the study.

\subsection{Participant Background Questionnaire}
\label{app:participant_background}
\textit{Please select or fill in the option that best describes you.}

\begin{enumerate}[leftmargin=*, label=(\arabic*), topsep=0pt]
    \item \textbf{Participant ID:} \_\_\_\_\_\_\_\_\_\_\_\_\_\_\_\_
    \item \textbf{Name:} \_\_\_\_\_\_\_\_\_\_\_\_\_\_\_\_
    \item \textbf{Gender:} \singlechoice{} Male \quad \singlechoice{} Female \quad \singlechoice{} Other / Prefer not to say
    \item \textbf{Age range:} \singlechoice{} <30 \quad \singlechoice{} 30-39 \quad \singlechoice{} 40-49 \quad \singlechoice{} 50+
    \item \textbf{What is your clinical specialty?}
        \begin{itemize}[label={}]
            \item \singlechoice{} A. Nephrology
            \item \singlechoice{} B. Obstetrics
            \item \singlechoice{} C. Other (Please specify): \_\_\_\_\_\_\_\_\_\_\_\_\_\_\_\_
        \end{itemize}
    \item \textbf{How many years have you been in clinical practice?}
        \begin{itemize}[label={}]
            \item \singlechoice{} A. Less than 1 year or in internship/rotation
            \item \singlechoice{} B. 1-5 years
            \item \singlechoice{} C. 6-10 years
            \item \singlechoice{} D. 11-15 years
            \item \singlechoice{} E. 16-20 years
            \item \singlechoice{} F. More than 20 years
        \end{itemize}
    \item \textbf{What is your current role or professional title?}
        \begin{itemize}[label={}]
            \item \singlechoice{} A. Medical Student (currently enrolled)
            \item \singlechoice{} B. Intern / Resident in Training
            \item \singlechoice{} C. Resident Physician
            \item \singlechoice{} D. Attending Physician
            \item \singlechoice{} E. Associate Chief Physician
            \item \singlechoice{} F. Chief Physician
            \item \singlechoice{} G. Other (Please specify): \_\_\_\_\_\_\_\_\_\_\_\_\_\_\_\_
        \end{itemize}
    \item \textbf{How familiar are you with AI-powered clinical decision support systems?}
        \begin{itemize}[label={}]
            \item \singlechoice{} 1: Not familiar at all (I have not heard of this concept).
            \item \singlechoice{} 2: Slightly familiar (I have heard of it, but am not sure what it does).
            \item \singlechoice{} 3: Moderately familiar (I understand its general definition and purpose).
            \item \singlechoice{} 4: Very familiar (I have a deeper understanding of its applications and key technologies).
            \item \singlechoice{} 5: Expert (I have professional knowledge or practical experience in this field).
        \end{itemize}
    \item \textbf{Prior to this study, how often did you use AI-based decision support systems in your clinical work?}
        \begin{itemize}[label={}]
            \item \singlechoice{} A. Never
            \item \singlechoice{} B. Rarely (e.g., once or twice a month)
            \item \singlechoice{} C. Occasionally (e.g., a few times a month)
            \item \singlechoice{} D. Frequently (e.g., on most workdays)
        \end{itemize}
\end{enumerate}

\subsection{Case-specific Questionnaire}
\label{app:case_questionnaire}
\textit{For the patient case you just reviewed, please answer the following questions.}

\textbf{Part 1: Nephrology Cases}
\begin{enumerate}[leftmargin=*, label=(\arabic*), topsep=0pt]
    \item \textbf{What is your assessment of this patient's risk of mortality within the next year?}
    \begin{itemize}[label={}]
        \item \singlechoice{} A. 0-25\% (Low risk): Very unlikely to die within 1 year.
        \item \singlechoice{} B. 25-50\% (Low-medium risk): Generally stable, but with some health risks.
        \item \singlechoice{} C. 50-75\% (Medium-high risk): Significant health concerns that require attention.
        \item \singlechoice{} D. 75-100\% (High risk): Critical health indicators, high probability of death within 1 year.
    \end{itemize}
    \item \textbf{How confident are you in this assessment?} (1=Not confident at all, 5=Very confident) \\
    {\singlechoice{} 1 \quad \singlechoice{} 2 \quad \singlechoice{} 3 \quad \singlechoice{} 4 \quad \singlechoice{} 5}
    \item \textbf{Which features were most important for your assessment? (Select all that apply)}
    \begin{itemize}[label={}]
        \item \multichoice{} Albumin \quad \multichoice{} Diastolic BP \quad \multichoice{} Systolic BP \quad \multichoice{} Urea \quad \multichoice{} Weight
        \item \multichoice{} Serum Sodium \quad \multichoice{} Serum Chloride \quad \multichoice{} Hemoglobin \quad \multichoice{} Creatinine
        \item \multichoice{} Serum Phosphorus \quad \multichoice{} Serum Potassium \quad \multichoice{} Food Intake \quad \multichoice{} Serum Calcium
        \item \multichoice{} Bicarbonate \quad \multichoice{} hs-CRP \quad \multichoice{} Blood Glucose \quad \multichoice{} White Blood Cell Count
        \item \multichoice{} Other: \_\_\_\_\_\_\_\_\_\_
    \end{itemize}
\end{enumerate}

\textbf{Part 2: Obstetrics Cases}
\begin{enumerate}[leftmargin=*, label=(\arabic*), topsep=0pt]
    \item \textbf{What is your assessment of this patient's risk of spontaneous preterm birth (before 37 weeks)?}
    \begin{itemize}[label={}]
        \item \singlechoice{} A. 0-25\% (Low risk): Very likely to carry to term.
        \item \singlechoice{} B. 25-50\% (Low-medium risk): Some risk factors present, but preterm birth is not expected.
        \item \singlechoice{} C. 50-75\% (Medium-high risk): Significant concern for preterm birth, may require intervention.
        \item \singlechoice{} D. 75-100\% (High risk): Preterm birth is highly probable without intervention.
    \end{itemize}
    \item \textbf{How confident are you in this assessment?} (1=Not confident at all, 5=Very confident) \\
    {\singlechoice{} 1 \quad \singlechoice{} 2 \quad \singlechoice{} 3 \quad \singlechoice{} 4 \quad \singlechoice{} 5}
    \item \textbf{Which features were most important for your assessment? (Select all that apply)}
    \begin{itemize}[label={}]
        \item \multichoice{} Activated Partial Thromboplastin Time (APTT)
        \item \multichoice{} Prothrombin Time (PT) \quad \multichoice{} Thrombin Time (TT)
        \item \multichoice{} Fibrinogen (Fib)
        \item \multichoice{} Thyroid Stimulating Hormone (TSH) \quad \multichoice{} Thyroxine (T4)
        \item \multichoice{} Albumin (ALB) \quad \multichoice{} Total Protein (TP)
        \item \multichoice{} Potassium (K) \quad \multichoice{} Chloride (Cl) \quad \multichoice{} Glucose (GLU)
        \item \multichoice{} Red Blood Cell Count (RBC) \quad \multichoice{} Other: \_\_\_\_\_\_\_\_\_\_
    \end{itemize}
\end{enumerate}

\subsection{NASA-Task Load Index (NASA-TLX)}
\label{app:nasa_tlx}
\textit{This scale assesses the workload you experienced during the last block of tasks. Please place a mark on each scale that matches your experience.}

\begin{itemize}[leftmargin=*, topsep=0pt]
    \item \textbf{Mental Demand}: How much mental and perceptual activity was required (e.g., thinking, deciding, remembering, connecting patient data)? \\
    Low \rule{0.7\linewidth}{0.4pt} High
    \item \textbf{Physical Demand}: How much physical activity was required (e.g., clicking, scrolling, interacting with the interface)? \\
    Low \rule{0.7\linewidth}{0.4pt} High
    \item \textbf{Temporal Demand}: How much time pressure did you feel due to the rate or pace at which the tasks needed to be completed? \\
    Low \rule{0.7\linewidth}{0.4pt} High
    \item \textbf{Performance}: How successful do you think you were in accomplishing the goals of the diagnostic task? (Note the direction). \\
    Good \rule{0.7\linewidth}{0.4pt} Poor
    \item \textbf{Effort}: How hard did you have to work (mentally and physically) to accomplish your level of performance? \\
    Low \rule{0.7\linewidth}{0.4pt} High
    \item \textbf{Frustration}: How insecure, discouraged, irritated, stressed, and annoyed versus secure, gratified, content, and relaxed did you feel during the task? \\
    Low \rule{0.7\linewidth}{0.4pt} High
\end{itemize}

\subsection{System Usability Scale (SUS)}
\label{app:sus}

The specific items used in the System Usability Scale are listed in Table~\ref{tab:sus}.

\begin{table*}[!ht]
\caption{The system usability scale (SUS). For the following statements about the AICare system, please indicate your level of agreement.}
\label{tab:sus}
\resizebox{0.9\linewidth}{!}{%
\begin{tabular}{@{}clccccc@{}}
\toprule
\textbf{\#} & \textbf{Statement} & \multicolumn{1}{p{2cm}}{\centering Strongly \\ Disagree (1)} & \multicolumn{1}{p{1.5cm}}{\centering Disagree (2)} & \multicolumn{1}{p{1.5cm}}{\centering Neutral (3)} & \multicolumn{1}{p{1.5cm}}{\centering Agree (4)} & \multicolumn{1}{p{1.5cm}}{\centering Strongly \\ Agree (5)} \\
\midrule
1 & I think that I would like to use this system frequently in my clinical work. & \singlechoice{} & \singlechoice{} & \singlechoice{} & \singlechoice{} & \singlechoice{} \\
2 & I found the system unnecessarily complex. & \singlechoice{} & \singlechoice{} & \singlechoice{} & \singlechoice{} & \singlechoice{} \\
3 & I thought the system was easy to use. & \singlechoice{} & \singlechoice{} & \singlechoice{} & \singlechoice{} & \singlechoice{} \\
4 & I think I would need the support of a technical person to be able to use this system. & \singlechoice{} & \singlechoice{} & \singlechoice{} & \singlechoice{} & \singlechoice{} \\
5 & I found the various functions in this system (e.g., charts, lists) were well integrated. & \singlechoice{} & \singlechoice{} & \singlechoice{} & \singlechoice{} & \singlechoice{} \\
6 & I thought there was too much inconsistency in this system. & \singlechoice{} & \singlechoice{} & \singlechoice{} & \singlechoice{} & \singlechoice{} \\
7 & I would imagine that most clinicians would learn to use this system very quickly. & \singlechoice{} & \singlechoice{} & \singlechoice{} & \singlechoice{} & \singlechoice{} \\
8 & I found the system very cumbersome to use. & \singlechoice{} & \singlechoice{} & \singlechoice{} & \singlechoice{} & \singlechoice{} \\
9 & I felt very confident using the system. & \singlechoice{} & \singlechoice{} & \singlechoice{} & \singlechoice{} & \singlechoice{} \\
10 & I needed to learn a lot of things before I could get going with this system. & \singlechoice{} & \singlechoice{} & \singlechoice{} & \singlechoice{} & \singlechoice{} \\
\bottomrule
\end{tabular}
}
\end{table*}

\subsection{Trust in Automation Scale}
\label{app:trust}

The items comprising the Trust in Automation Scale are provided in Table~\ref{tab:trust}.

\begin{table*}[!ht]
\caption{The trust in automation scale. Please indicate your agreement with the following statements about the AICare system. (1=Strongly Disagree, 7=Strongly Agree). (R) indicates a reverse-scored item.}
\label{tab:trust}
\small
\resizebox{0.9\linewidth}{!}{%
\begin{tabular}{@{}p{0.01\linewidth}p{0.68\linewidth}ccccccc@{}}
\toprule
\textbf{\#} & \textbf{Statement} & \textbf{1} & \textbf{2} & \textbf{3} & \textbf{4} & \textbf{5} & \textbf{6} & \textbf{7} \\
\midrule
1 & The system is deceptive (e.g., it seems to hide important information or provide misleading results). (R) & \singlechoice{} & \singlechoice{} & \singlechoice{} & \singlechoice{} & \singlechoice{} & \singlechoice{} & \singlechoice{} \\
2 & The system's behavior is predictable (i.e., its analysis is consistent for similar patient cases). & \singlechoice{} & \singlechoice{} & \singlechoice{} & \singlechoice{} & \singlechoice{} & \singlechoice{} & \singlechoice{} \\
3 & The system is designed with the patient's best interest in mind. & \singlechoice{} & \singlechoice{} & \singlechoice{} & \singlechoice{} & \singlechoice{} & \singlechoice{} & \singlechoice{} \\
4 & I am unsure about the system's capabilities (i.e., when it is reliable and when it might fail). (R) & \singlechoice{} & \singlechoice{} & \singlechoice{} & \singlechoice{} & \singlechoice{} & \singlechoice{} & \singlechoice{} \\
5 & I am confident in the system. & \singlechoice{} & \singlechoice{} & \singlechoice{} & \singlechoice{} & \singlechoice{} & \singlechoice{} & \singlechoice{} \\
6 & The system has characteristics that could be harmful to my clinical decision-making. (R) & \singlechoice{} & \singlechoice{} & \singlechoice{} & \singlechoice{} & \singlechoice{} & \singlechoice{} & \singlechoice{} \\
7 & I am familiar with how the system works. & \singlechoice{} & \singlechoice{} & \singlechoice{} & \singlechoice{} & \singlechoice{} & \singlechoice{} & \singlechoice{} \\
8 & I can trust the system. & \singlechoice{} & \singlechoice{} & \singlechoice{} & \singlechoice{} & \singlechoice{} & \singlechoice{} & \singlechoice{} \\
9 & I am unfamiliar with how the system works. (R) & \singlechoice{} & \singlechoice{} & \singlechoice{} & \singlechoice{} & \singlechoice{} & \singlechoice{} & \singlechoice{} \\
10 & The system's explanations are a faithful representation of its reasoning (i.e., it honestly reflects its underlying data and logic). & \singlechoice{} & \singlechoice{} & \singlechoice{} & \singlechoice{} & \singlechoice{} & \singlechoice{} & \singlechoice{} \\
11 & I distrust the system. (R) & \singlechoice{} & \singlechoice{} & \singlechoice{} & \singlechoice{} & \singlechoice{} & \singlechoice{} & \singlechoice{} \\
12 & The system is reliable (i.e., its performance is stable and its findings are reproducible). & \singlechoice{} & \singlechoice{} & \singlechoice{} & \singlechoice{} & \singlechoice{} & \singlechoice{} & \singlechoice{} \\
\bottomrule
\end{tabular}
}
\end{table*}

\subsection{AICare Feature Feedback}
\label{app:feature_feedback}
\textit{Please rate the usefulness of each AICare feature for your diagnostic process.} (1=Not useful at all, 5=Extremely useful). The feedback form structure is shown in Table~\ref{tab:feature_feedback}.

\begin{table*}[!ht]
    \centering
    \caption{Participant rating of the usefulness of each AICare feature.}
    \label{tab:feature_feedback}
% \resizebox{1.0\columnwidth}{!}{%
\begin{tabular}{p{8cm}ccccc}
\toprule
\textbf{Feature} & \textbf{1} & \textbf{2} & \textbf{3} & \textbf{4} & \textbf{5} \\
\midrule
Dynamic risk trajectory visualization & \singlechoice{} & \singlechoice{} & \singlechoice{} & \singlechoice{} & \singlechoice{} \\
Interactive list of critical risk factors & \singlechoice{} & \singlechoice{} & \singlechoice{} & \singlechoice{} & \singlechoice{} \\
Population-level indicator analysis visualization & \singlechoice{} & \singlechoice{} & \singlechoice{} & \singlechoice{} & \singlechoice{} \\
LLM-driven diagnostic recommendation & \singlechoice{} & \singlechoice{} & \singlechoice{} & \singlechoice{} & \singlechoice{} \\
\bottomrule
\end{tabular}
% }
\end{table*}

\subsection{Semi-structured Interview Guide}
\label{app:interview}
\textit{Researcher's Guide: The goal is to understand the ``why'' behind the quantitative results. Use these questions as a guide, but feel free to ask follow-up questions to probe deeper into interesting comments. Start by making the participant feel comfortable.}

\begin{enumerate}[leftmargin=*, label=(\arabic*), topsep=0pt]
    \item \textbf{On Usability and Workflow Integration:}
    \begin{itemize}[leftmargin=*, topsep=0pt]
        \item How easy or difficult was it to understand the information presented by AICare? Was any part of the interface confusing or overwhelming?
        \item Imagine if your department introduced this tool tomorrow. How do you think it would integrate into your daily work?
        \item Do you think it could help you improve diagnostic efficiency and accuracy?
    \end{itemize}
    \item \textbf{On Trust, Interpretability, and Clinical Reasoning:}
    \begin{itemize}[leftmargin=*, topsep=0pt]
        \item Let's talk about trust. When you were using AICare, were there any specific moments or features that made you start to trust its predictions? Conversely, were there times you felt skeptical?
        \item I would like to talk specifically about a few of the explanatory features:
        \begin{itemize}[leftmargin=*, topsep=0pt]
            \item How did the \textbf{dynamic risk trajectory visualization} influence your understanding of the patient's condition? Did seeing the historical and predicted trends help you build a holistic view of the patient's situation?
            \item What about the \textbf{interactive list of critical risk factors}? Can you recall an instance where you hovered over or clicked on an indicator to see its trend? How did that action affect your trust in the AI's reasoning?
            \item Finally, what are your thoughts on the \textbf{LLM-driven diagnostic recommendation}? Did you find them useful? Did they feel consistent with the other explanatory features?
            \item The system also showed \textbf{population-level indicator analysis visualization}. Did you find this feature to be a useful supplement to the AI's explanation, or was it redundant?
        \end{itemize}
        \item During your use, was there a time when your own clinical judgment and the AI's prediction were inconsistent? Can you describe that situation and how you handled the conflict? Did the system's explanatory features help you understand the difference?
    \end{itemize}
    \item \textbf{On Clinical Value and Impact:}
    \begin{itemize}[leftmargin=*, topsep=0pt]
        \item Did the system ever provide you with a new insight or make you notice a risk factor you might have otherwise overlooked? Can you give an example?
    \end{itemize}
    \item \textbf{Closing Questions:}
    \begin{itemize}[leftmargin=*, topsep=0pt]
        \item Is there anything else you would like to share about your experience today?
    \end{itemize}
\end{enumerate}

\end{document}